\documentclass[lettersize,journal]{IEEEtran}

\usepackage{amsmath,amsfonts}
\usepackage{amssymb}
\usepackage{amsthm}
\usepackage{algorithmic}
\usepackage{algorithm}
\usepackage{array}
\usepackage[caption=false,font=normalsize,labelfont=sf,textfont=sf]{subfig}
\usepackage{textcomp}
\usepackage{stfloats}
\usepackage{multirow}
\usepackage{url}
\usepackage{xurl}
\usepackage{verbatim}
\usepackage{graphicx}
\usepackage{cite}
\usepackage{booktabs}
\usepackage{capt-of}
\theoremstyle{plain}
\newtheorem{theorem}{Theorem}

\newtheorem{proposition}{Proposition}

\theoremstyle{definition}

\begin{document}

\title{Absolute-Time Network Calculus for Deterministic Performance Analysis of LEO Satellite Networks with Time-Aware Shaping}
\author{Yaohua~Sun,
        Huiqi~Feng,
        Xin'ao~Feng,
        and~Mugen~Peng,~\IEEEmembership{Fellow,~IEEE}%
\thanks{Yaohua~Sun, Huiqi~Feng, Xin'ao~Feng, and Mugen~Peng are with the Beijing Key Laboratory of Convergent Communications and Networking Technologies in LEO Satellite Systems and the State Key Laboratory of Networking and Switching Technology, Beijing University of Posts and Telecommunications, Beijing 100876, China (e-mail: sunyaohua@bupt.edu.cn; fhq@bupt.edu.cn; fengxinao@bupt.edu.cn; pmg@bupt.edu.cn).}
\thanks{This work has been submitted to the IEEE for possible publication. Copyright may be transferred without notice, after which this version may no longer be accessible.}
}
\maketitle

\begin{abstract}
Low Earth orbit (LEO) satellite networks are expected to support time-sensitive services requiring deterministic performance guarantees. To support these guarantees, deterministic network calculus provides worst-case performance bounds for Time-Aware Shaping (TAS), which schedules packet transmissions through periodic Gate Control Lists (GCLs). However, time-varying inter-satellite link (ISL) propagation delays shift downstream packet arrivals relative to transmission windows, making valid and tight multihop bounds difficult to derive. To address this challenge, we establish an absolute-time network calculus framework for deterministic performance analysis of TAS in LEO satellite networks. Specifically, we first characterize packet arrivals and GCL service over absolute time intervals to derive single-port performance bounds and per-flow output bounds. To extend this analysis to multihop transmission, we transform upstream output bounds into downstream arrival bounds under time-varying ISL propagation delays. Moreover, packet ordering is used to exclude packets that cannot affect the transmission of the target packet, thereby tightening performance bounds. Numerical results show that, compared with three network-calculus baselines, the proposed method achieves improvements of 26.0\%--51.9\%, 68.8\%--86.7\%, 2.6\%--24.1\%, and 37.5\%--73.8\% in delay, jitter, throughput, and the sum of per-port backlog bounds, respectively. These bounds closely match packet-level simulation results, particularly for throughput and backlog.
\end{abstract}


\section{Introduction}
\label{sec:introduction}

Low Earth orbit (LEO) mega-constellations have expanded rapidly and are emerging as an important component of wide-area connectivity infrastructure~\cite{zhu2025beam}. Equipped with optical inter-satellite links (ISLs), satellites can forward traffic on board and route it over multiple inter-satellite hops, allowing a large constellation to operate as a space network \cite{feng2024routing}. Beyond broadband access and data backhaul, LEO networks are also expected to support time-sensitive (TS) services, including onboard collaborative control, autonomous driving \cite{li2026safvin}, and telemedicine. Such services require not only connectivity \cite{sun2026handover} and throughput but also deterministic bounds on end-to-end delay and jitter.

To provide such guarantees, Time-Sensitive Networking (TSN) extends bridged Ethernet to support deterministic packet transport \cite{lobello2019perspective}. Although TSN has primarily been developed and applied in terrestrial networks, a contribution to IEEE 802.1 has proposed developing a TSN profile for terrestrial--space and multi-orbit satellite-to-satellite communications \cite{zein2025spaceprofile}. TSN includes both asynchronous and synchronous traffic control mechanisms, with the latter coordinating transmission opportunities across nodes using a common time reference. The applicability of such coordination to LEO satellite networks is supported by measurements of Starlink, which reported globally synchronized changes in latency and throughput at regular intervals, suggesting periodic coordination of network resources \cite{mohan2024starlink}. These findings motivate the study of synchronous TSN mechanisms in LEO satellite networks.

Among these mechanisms, Time-Aware Shaping (TAS) \cite{ieee8021q2022} has attracted considerable research attention because its periodically repeated Gate Control List (GCL) enables precise scheduling of transmission windows, thereby supporting deterministic performance \cite{ieee80224tsn2019,zhao2022shapers}. However, designing TAS configurations for multihop traffic is challenging because GCL transmission windows must be coordinated across successive nodes to satisfy the timing requirements of multiple TS flows \cite{craciunas2016scheduling}. Consequently, worst-case analysis is needed to evaluate candidate configurations and verify whether they can guarantee the strict delay and jitter requirements of TS traffic \cite{rfc9320}. The need for analytical verification is also reflected in a NASA presentation, which calls for worst-case traversal-time calculations to verify timing requirements \cite{donnelly2022nasa}. Overall, an analytical framework is needed to derive deterministic performance bounds for TS traffic under given TAS scheduling configurations in LEO networks.

Deterministic network calculus provides a systematic framework for establishing worst-case performance guarantees, which has been used to verify delay and buffer bounds in Avionics Full-Duplex Switched Ethernet (AFDX) networks deployed on the Airbus A380 and Boeing 787 \cite{pasquier2015afdx}. In this framework, arrival curves bound the amount of traffic that can arrive within a given time interval, whereas service curves provide lower bounds on the service available to backlogged traffic \cite{leboudec2001networkcalculus}. By combining these bounds, worst-case delay and backlog bounds can be derived for individual nodes and multihop paths. These bounds can be used not only to verify whether a given configuration guarantees the required performance of TS traffic but also to provide quantitative guidance for TAS configuration design by comparing different configurations. However, in LEO satellite networks, ISL propagation delays vary with satellite motion \cite{jiang2023qoe}, changing packet arrival times at downstream nodes and posing additional challenges for multihop performance analysis. Therefore, a systematic analysis is needed to account for these variations and derive tighter deterministic performance bounds for multihop TAS in LEO satellite networks.

\subsection{Related Work}

Network calculus has been widely used to derive worst-case performance bounds for TAS in terrestrial TSN networks. Zhao et al. constructed service curves based on the guaranteed service available within GCL windows and derived end-to-end delay bounds under given GCL configurations \cite{zhao2018qbv}. However, their analysis does not account for the relative positions of transmission windows at consecutive nodes, which may lead to conservative multihop delay bounds. To account for timing relationships across successive TAS nodes, Zhao et al. incorporated GCL window offsets and constraints imposed by upstream transmission windows to derive tighter end-to-end delay bounds \cite{zhao2021offsets}. Hu et al. likewise incorporated inter-node GCL window offsets into service analysis and refined traffic arrival curves using finite link transmission rates \cite{hu2025upperbound}. However, these analyses still bound traffic arrivals based only on interval duration. Consequently, the resulting arrival bounds may be combined with GCL service conditions that cannot occur at the same time. The importance of preserving timing information has also been examined in more general network calculus studies. Daniel-Cavalcante and Santos-Mendes developed a time-varying network calculus framework using time-dependent arrival and service constraints \cite{daniel2013timevarying}. Falk et al. further showed that preserving the timing of periodically interrupted service can provide tighter delay and backlog bounds \cite{falk2019intermittence}. While these studies provide a theoretical basis for analyzing time-dependent traffic and service, multihop TAS analysis involves additional timing relationships across successive nodes. Specifically, packet departures constrained by upstream GCL windows determine when packets arrive at subsequent nodes, where their waiting times depend on their arrival times relative to local transmission windows. Existing analyses still do not fully capture these timing relationships, which may lead to conservative end-to-end performance bounds.

In LEO satellite networks, time-varying ISL propagation introduces an additional challenge for multihop performance analysis. Existing studies have often treated propagation delay separately from downstream traffic arrivals. For example, Wang et al. improved the service-curve model for TAS to derive tighter end-to-end delay bounds and treated propagation delay as an additional link delay \cite{wang2025leodelay}, while Zhang et al. employed martingale theory to derive probabilistic delay and backlog bounds for multihop heterogeneous satellite networks without considering propagation delay in the queueing analysis \cite{zhang2025martingale}. As a result, these analyses do not capture how changes in propagation delay affect downstream traffic arrivals, which may compromise the validity or tightness of subsequent queueing delay and backlog bounds. The impact of timing changes on traffic arrival bounds has also been examined in network calculus. Thomas and Le Boudec derived arrival-curve transformations between nonideal clocks by bounding changes in interval duration under time mappings \cite{thomas2020clocks}. In LEO satellite networks, Chen et al. related downstream arrivals to upstream departures through time-varying propagation delays \cite{chen2026martingale}. However, their analysis assumes constant propagation delay within each time slot and thus does not capture the effects of propagation variations within a slot. In practice, continuous variations in propagation delay can compress or expand the intervals between packet arrivals at successive nodes. When these changes in interval length are ignored, applying upstream output bounds to downstream arrival intervals may underestimate the number of arriving packets or produce unnecessarily loose bounds.

Furthermore, conservatism in multihop performance analysis can accumulate as traffic bounds are passed from one node to the next. Several studies have used more detailed traffic information and relationships among flows to obtain tighter bounds. For periodic flows commonly found in TSN, Tabatabaee et al. developed finite-horizon total flow analysis (FH-TFA), which enables the use of detailed periodic arrival curves by restricting the analysis to finite time intervals \cite{tabatabaee2023fhtfa}. To better capture FIFO ordering constraints, Bouillard and Stea combined arrival and service constraints with FIFO input-output relationships to compute exact worst-case delays for a class of feedforward networks \cite{bouillard2015exact}. Geyer et al. used flow prolongation to better account for interactions among flows in multihop FIFO networks and reduce the conservatism caused by flow decomposition \cite{geyer2023flow}. However, such flow-level traffic bounds do not necessarily identify which packets still need to be served before a given target packet under FIFO scheduling. Consequently, the analysis may include packets that have already completed serialization before the target packet arrives or that arrive after it. This issue is particularly important in LEO satellite networks, where time-varying ISL propagation delays can shift packet arrival times relative to competing traffic and GCL windows. Under TAS, even a small overestimation of the number of competing packets may lead to an estimated transmission time in a later GCL window, resulting in conservative end-to-end performance bounds.

Overall, existing studies have improved traffic and service analysis for TAS, examined time-varying propagation delays in LEO satellite networks, and developed methods to reduce conservatism in multihop performance analysis. However, several challenges remain when these aspects are considered together in multihop TAS analysis for LEO satellite networks. Specifically, packet departure times are constrained by upstream GCL windows, while variations in ISL propagation delay shift downstream arrival times and can compress or expand packet arrival intervals. These changes affect both the timing of packet arrivals relative to downstream GCL windows and the competition among packets for service, with their effects extending to subsequent hops. These interactions make it difficult to derive valid and tight deterministic end-to-end performance bounds.

\subsection{Contribution}

To address these challenges, this paper establishes an absolute-time network calculus framework for deriving deterministic performance bounds for TAS in LEO satellite networks with time-varying ISL propagation delays. The main contributions are summarized as follows.

\begin{itemize}

\item We develop an absolute-time network calculus method for analyzing TAS output ports in LEO satellite networks. In this method, packet arrivals and GCL service are characterized over absolute time intervals to preserve the timing relationships between packet arrivals and GCL transmission windows. Based on these characterizations, we derive single-port performance bounds and per-flow output bounds for downstream arrival analysis.

\item Since time-varying ISL propagation delays can compress or expand packet arrival intervals during multihop transmission, we establish an arrival bound transformation that maps each downstream arrival interval back to its corresponding upstream departure interval. This transformation ensures valid downstream arrival bounds without introducing unnecessary conservatism.

\item To further tighten the performance bounds, we exploit packet ordering at each TAS output port to exclude packets that arrive after the target packet or have departed before its arrival. The remaining packet count is then combined with the single-port backlog bound to limit the number of packets considered in the performance analysis. Finally, we use the single-port analysis, arrival bound transformation, and packet-ordering refinement to establish deterministic bounds on end-to-end delay, jitter, and throughput, as well as per-port backlog.

\item Numerical results show that, compared with three network-calculus baselines, the proposed method achieves improvements of 26.0\%--51.9\%, 68.8\%--86.7\%, 2.6\%--24.1\%, and 37.5\%--73.8\% in delay, jitter, throughput, and the sum of per-port backlog bounds, respectively. The resulting bounds closely match packet-level simulation results, particularly for throughput and backlog. Additional experiments confirm the validity and tightness of the transformed arrival bounds and the effectiveness of packet-ordering refinement. Moreover, sensitivity analysis demonstrates consistent advantages over the baselines across varying GCL window lengths and cross-traffic loads.
\end{itemize}

\begin{figure*}[t]
    \centering
    \includegraphics[width=\textwidth]{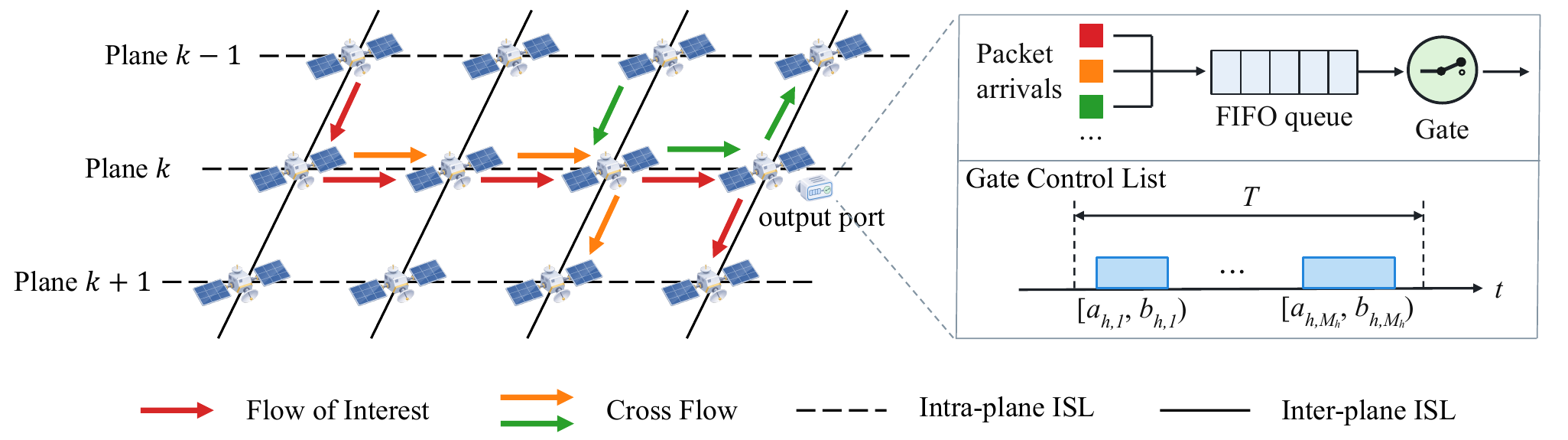}
    \caption{Illustration of the system model.}
    \label{fig:system-model}
\end{figure*}

\begin{table*}[t]
\caption{Main Notations}
\label{tab:main-notations}
\centering
\footnotesize
\setlength{\tabcolsep}{3pt}
\renewcommand{\arraystretch}{1.1}
\begin{tabular}{
@{}
>{\raggedright\arraybackslash}p{0.11\textwidth}
>{\raggedright\arraybackslash}p{0.36\textwidth}
@{\hspace{0.02\textwidth}}
>{\raggedright\arraybackslash}p{0.11\textwidth}
>{\raggedright\arraybackslash}p{0.36\textwidth}
@{}
}
\toprule
\textbf{Symbol} & \textbf{Meaning}
& \textbf{Symbol} & \textbf{Meaning} \\
\midrule

$f^\star$
& Flow of interest
& $H$
& Number of output ports on the routing path \\
\specialrule{0.25pt}{1pt}{1pt}

$p_h$
& $h$-th output port on the routing path
& $v_h$
& Satellite at which output port $p_h$ is located \\
\specialrule{0.25pt}{1pt}{1pt}

$\mathcal P_{f^\star}$
& Routing path of $f^\star$
& $e_h$
& Directed ISL from $v_h$ to the next satellite $v_{h+1}$ \\
\specialrule{0.25pt}{1pt}{1pt}

$[0,R)$
& Release interval of evaluated packets of $f^\star$
& $\mathcal J$
& Evaluation interval \\
\specialrule{0.25pt}{1pt}{1pt}

$\phi_h(s)$
& Arrival time at the next satellite $v_{h+1}$ for a packet completing serialization at $p_h$ at time $s$
& $\mathcal F$
& Set of periodic TS flows traversing at least one output port on $\mathcal P_{f^\star}$ \\
\specialrule{0.25pt}{1pt}{1pt}

$\delta_h(s)$
& Propagation delay of ISL $e_h$ at time $s$
& $\mathcal F_h$
& Set of flows traversing $p_h$ \\
\specialrule{0.25pt}{1pt}{1pt}

$a_{h,f,n},\,C_{h,f,n}$
& Arrival and serialization completion times of packet $(f,n)$ at $p_h$
& $\mathcal N_{f^\star}$
& Indices of packets of $f^\star$ released during $[0,R)$ \\
\specialrule{0.25pt}{1pt}{1pt}

$\gamma_h$
& GCL phase at $p_h$
& $\mathcal O_h$
& Set of gate-open times at $p_h$ \\
\specialrule{0.25pt}{1pt}{1pt}

$q_h$
& Serialization time of a packet at $p_h$
& $d_{f^\star,n}^{\mathrm{e2e}}$
& End-to-end delay of packet $n$ of $f^\star$ \\
\specialrule{0.25pt}{1pt}{1pt}

$J_{f^\star}$
& End-to-end jitter of $f^\star$
& $W$
& Throughput window length \\
\specialrule{0.25pt}{1pt}{1pt}

$\Theta_{f^\star}(W)$
& Throughput of $f^\star$ over $[0,W]$
& $Q_h(t)$
& Backlog at $p_h$ at time $t$, in packets \\
\specialrule{0.25pt}{1pt}{1pt}

$D_{f^\star}^{\mathrm U},\,J_{f^\star}^{\mathrm U}$
& Upper bounds on end-to-end delay and jitter of $f^\star$
& $\Theta_{f^\star}^{\mathrm L}(W)$
& Lower bound on throughput of $f^\star$ over $[0,W]$ \\
\specialrule{0.25pt}{1pt}{1pt}

$B_h^{\mathrm U}$
& Upper bound on backlog at $p_h$ over $\mathcal J$
& $\Gamma_{\Sigma,h}^{\mathrm{in}}(u,x)$
& Upper bound on total arrivals at $p_h$ during $[u,u+x]$ \\
\specialrule{0.25pt}{1pt}{1pt}

$\Gamma_{f,h}^{\mathrm{in}}(u,x)$
& Upper bound on arrivals of flow $f$ at $p_h$ during $[u,u+x]$
& $c_{h,m}(u)$
& Serialization completion time of the $m$-th packet in the reference queue backlogged from time $u$ \\
\specialrule{0.25pt}{1pt}{1pt}

$S_h(u,v)$
& Number of packets completing serialization in the reference queue during $[u,v]$
& $d_h^{\mathrm U}(t)$
& Delay upper bound at $p_h$ for packets arriving at time $t$ \\
\specialrule{0.25pt}{1pt}{1pt}

$\bar d_h$
& Uniform packet delay upper bound at $p_h$
& $\Gamma_{f,h}^{\mathrm{out}}(s,t-s)$
& Upper bound on departures of flow $f$ at $p_h$ during $[s,t]$ \\

\bottomrule
\end{tabular}
\end{table*}

\section{System Model}
\label{sec:model}

In this section, we model the transmission of periodic TS traffic in a LEO satellite network, where TAS controls transmission opportunities at each satellite output port and ISL propagation varies with satellite motion, as illustrated in Fig.~\ref{fig:system-model}. This model provides the basis for deriving deterministic performance bounds, with the main notations used in this paper summarized in Table~\ref{tab:main-notations}.

\subsection{LEO Satellite Network Model}
\label{subsec:leo-network}

We consider a LEO satellite network~\cite{yuan2024placement} formed by a Walker Delta constellation with $N_P$ orbital planes and $N_S$ uniformly spaced satellites per plane.
All satellites have orbital altitude $h_{\mathrm{orb}}$ and inclination $\iota$~\cite{zhu2024timing}. Each satellite maintains four bidirectional optical ISLs, including two intra-plane links to adjacent satellites in the same plane and two inter-plane links to corresponding satellites in adjacent orbital planes \cite{feng2026jrs}. The mesh topology is represented by a directed graph $\mathcal G=(\mathcal V,\mathcal E)$ \cite{lu2026lecmor}, where $\mathcal V$ denotes the set of satellites and each bidirectional ISL contributes two oppositely directed edges to $\mathcal E$.

Motivated by the periodic network reconfigurations observed in Starlink \cite{mohan2024starlink}, we consider a given routing and TAS configuration, with all event times referenced to an ideal common network clock. Under this configuration, satellite motion continuously changes the distances of inter-plane ISLs and their corresponding propagation delays \cite{ren2026meta}. Let $f^\star$ denote the flow of interest, whose routing path consists of $H\ge1$ transmitting output ports. The $h$-th output port is denoted by $p_h$ and is located at satellite $v_h$. The routing path is represented by
\begin{equation}
\mathcal P_{f^\star}=(p_1,\ldots,p_H).
\label{eq:routing-path}
\end{equation}
For $h=1,\ldots,H$, the directed ISL $e_h=(v_h,v_{h+1})\in\mathcal E$ connects $v_h$ to the next satellite, with $v_{H+1}$ being the destination satellite of $f^\star$.

We consider packets of $f^\star$ released during $[0,R)$, where $R>0$. Since their end-to-end transmission may extend beyond $R$, we define an evaluation interval
\begin{equation}
\mathcal J=[0,E],
\qquad E\ge R.
\label{eq:evaluation-interval}
\end{equation}
The routing paths and TAS configurations remain fixed throughout $\mathcal J$. We assume that the associated ISLs remain available throughout this interval and that the routing paths do not include inter-plane seam ISLs.

\subsection{ISL Propagation Delay Model}
\label{subsec:isl-propagation}

For an ISL $e\in\mathcal E$, a packet completes serialization at time $s$ when its last bit is transmitted onto the link. Its arrival time at the next satellite is then given by
\begin{equation}
\phi_e(s)=s+\delta_e(s),
\label{eq:propagation-map}
\end{equation}
where $\delta_e(s)\ge0$ is the propagation delay of ISL $e$. For an intra-plane ISL, the propagation delay is constant and given by
\begin{equation}
\delta_e=
\frac{2r_o}{c}\sin\left(\frac{\Delta\Phi}{2}\right),
\label{eq:intraplane-delay}
\end{equation}
where $r_o=R_\oplus+h_{\mathrm{orb}}$ is the orbital radius, $R_\oplus$ is the Earth's radius, $c$ is the speed of light, and $\Delta\Phi=2\pi/N_S$ is the angular separation between adjacent satellites in the same orbital plane.

For adjacent non-seam inter-plane ISLs, we index the orbital planes from $1$ to $N_P$ in increasing order of right ascension of the ascending node. Each such ISL connects corresponding satellites in planes $k$ and $k+1$, where $k=1,\ldots,N_P-1$, and its propagation delay is given by \cite{chen2024shortest}
\begin{equation}
\delta_e(s)=\frac{\sqrt{2}r_o}{c}\sqrt{A+B\cos\!\left(2u_e(s)+\eta_e\Delta f\right)},
\label{eq:interplane-delay}
\end{equation}
where $\eta_e=1$ for transmission from plane $k$ to plane $k+1$ and $\eta_e=-1$ for the reverse direction, and $u_e(s)=u_{e,0}+n_os$ is the argument of latitude of the transmitting satellite. Here, $u_{e,0}$ is the argument of latitude at time $0$, and $n_o=\sqrt{\mu/r_o^3}$ is the orbital angular speed of the satellites, where $\mu$ is the gravitational parameter of Earth. The angular separation between adjacent orbital planes and the Walker phasing offset are given by
\begin{equation}
\Delta\Omega=\frac{2\pi}{N_P},
\qquad
\Delta f=\frac{2\pi F}{N_PN_S},
\label{eq:walker-spacing}
\end{equation}
where $F$ is the Walker phasing parameter. The coefficients $A$ and $B$ are given by
\begin{align}
A={}&
1-
\left[
\cos^2\!\left(\frac{\Delta\Omega}{2}\right)
-\cos^2\!\iota\,
\sin^2\!\left(\frac{\Delta\Omega}{2}\right)
\right]\cos\Delta f
\nonumber\\
&+\cos\iota\,\sin\Delta\Omega\,\sin\Delta f,
\label{eq:interplane-A}\\
B={}&
\sin^2\!\iota\,
\sin^2\!\left(\frac{\Delta\Omega}{2}\right).
\label{eq:interplane-B}
\end{align}

The packet arrival time $\phi_e(s)$ is continuous and strictly increasing with $s$ over $\mathcal J$ for every ISL $e\in\mathcal E$, so packet order is preserved during propagation. For notational simplicity, we write $\delta_h(s)=\delta_{e_h}(s)$ and $\phi_h(s)=\phi_{e_h}(s)$ for the ISL $e_h$ following output port $p_h$.

\subsection{Traffic Model}
\label{subsec:traffic-model}

Let $\mathcal F$ denote the set of periodic TS flows that traverse at least one output port on $\mathcal P_{f^\star}$, including the flow of interest $f^\star$. All flows in $\mathcal F$ have the same priority, and their packets have the same length $\ell>0$. The other flows in $\mathcal F_{\mathrm{cross}}=\mathcal F\setminus\{f^\star\}$ are referred to as cross flows and compete with $f^\star$ for transmission at shared output ports. For each $f\in\mathcal F$, let $p_{s_f}$ and $p_{t_f}$ denote the first and last output ports on $\mathcal P_{f^\star}$ traversed by the flow, respectively, with $1\le s_f\le t_f\le H$. For the flow of interest $f^\star$, we have $s_{f^\star}=1$ and $t_{f^\star}=H$. Each cross flow $f\in\mathcal F_{\mathrm{cross}}$ traverses consecutive output ports on $\mathcal P_{f^\star}$ and may enter or leave the path at an intermediate output port. The set of flows traversing output port $p_h$ is $\mathcal F_h=\{f\in\mathcal F:s_f\le h\le t_f\}$. We assume that immediately before time $0$, all queues are empty, no packet is being serialized, and no packet is in flight on any ISL. Each flow $f\in\mathcal F$ generates packets periodically at its source with period $P_f>0$ and release offset $\theta_f\in[0,P_f)$ relative to time $0$. The release time of packet $n$ is
\begin{equation}
r_{f,n}^{\mathrm{src}}
=
\theta_f+nP_f,
\qquad
n\in\mathbb Z_{\ge0}.
\label{eq:source-release}
\end{equation}

Let $a_{h,f,n}$ and $C_{h,f,n}$ denote the arrival time and serialization completion time of packet $n$ of flow $f$ at output port $p_h$, respectively. Packets of $f^\star$ are released directly into the queue at $p_1$, so $a_{1,f^\star,n}=r_{f^\star,n}^{\mathrm{src}}$. Cross flows may enter $\mathcal P_{f^\star}$ directly from their sources or through upstream satellite output ports and ISLs. For a cross flow $f$ entering directly from its source, $a_{s_f,f,n}=r_{f,n}^{\mathrm{src}}$. Otherwise, the arrival times depend on upstream TAS service and ISL propagation and are not necessarily periodic. The upstream TAS analysis includes all flows competing at these ports, including flows that do not traverse $\mathcal P_{f^\star}$. We neglect onboard processing and internal forwarding delays, so a packet arriving at an intermediate satellite enters the next output-port queue on its path at the same time. Accordingly, for $f\in\mathcal F$ and $s_f\le h<t_f$, the arrival time at the next output port satisfies $a_{h+1,f,n}=\phi_h(C_{h,f,n})$. The packets of $f^\star$ released during $[0,R)$ are indexed by
\begin{equation}
\mathcal N_{f^\star}
=
\left\{
n\in\mathbb Z_{\ge0}:
a_{1,f^\star,n}\in[0,R)
\right\}.
\label{eq:packet-index-set}
\end{equation}

We evaluate the end-to-end delay and jitter of the packets in $\mathcal N_{f^\star}$. All packets released during $\mathcal J$ are taken into account, as they may compete for transmission and contribute to backlog at the output ports.

\subsection{TAS Mechanism}
\label{subsec:tas-mechanism}

At output port $p_h$, TAS controls packet transmission from the FIFO queue over ISL $e_h$ according to its GCL. The GCL repeats with period $T>0$, shared by all output ports, and specifies $M_h\ge1$ gate-open intervals $[a_{h,j},b_{h,j})$, $j=1,\ldots,M_h$, per period, where
\begin{equation}
0\le a_{h,1}<b_{h,1}
\le a_{h,2}<b_{h,2}
\le\cdots\le
a_{h,M_h}<b_{h,M_h}\le T.
\label{eq:gcl-window-order}
\end{equation}

Let $\gamma_h\in[0,T)$ denote the GCL phase at $p_h$, defined as its time offset relative to the common network clock. The gate-open times over all periods are given by
\begin{equation}
\mathcal O_h
=
\bigcup_{m\in\mathbb Z}
\bigcup_{j=1}^{M_h}
\left[
\gamma_h+mT+a_{h,j},
\gamma_h+mT+b_{h,j}
\right).
\label{eq:gate-open-set}
\end{equation}
During these gate-open times, the FIFO queue has exclusive access to ISL $e_h$, and packets are transmitted at rate $C_h>0$. Since all packets have length $\ell$, their serialization time at $p_h$ is $q_h=\ell/C_h$.

We assume non-preemptive packet transmission at each output port, so serialization cannot be interrupted once it begins. A packet can start serialization at time $s$ only if its entire serialization interval lies within the gate-open times, as expressed by
\begin{equation}
[s,s+q_h)\subseteq\mathcal O_h.
\label{eq:tas-start-condition}
\end{equation}
The head-of-line packet starts serialization at the earliest time permitted by the GCL after it arrives and all preceding packets have completed serialization. If multiple packets arrive simultaneously, ties may be resolved arbitrarily.

\section{Analysis Objectives and Preliminaries}
\label{sec:analysis-foundations}

In this section, we define the performance metrics and deterministic bounds considered in this paper, and present the network calculus preliminaries used in the analysis.

\subsection{Analysis Objectives}
\label{subsec:analysis-objectives}

Given the traffic flows, routing path, and GCL configurations, we consider deterministic bounds on the end-to-end delay, jitter, and throughput of the flow of interest $f^\star$ and the backlog at each output port along its path.

For each packet $n\in\mathcal N_{f^\star}$ of $f^\star$ released during $[0,R)$, the end-to-end delay is
\begin{equation}
d^{\mathrm{e2e}}_{f^\star,n}
=
t^{\mathrm{dst}}_{f^\star,n}
-
a_{1,f^\star,n},
\label{eq:end-to-end-delay}
\end{equation}
where $t^{\mathrm{dst}}_{f^\star,n}=\phi_H\!\left(C_{H,f^\star,n}\right)$ denotes the reception time of packet $n$ at the destination. For nonempty $\mathcal N_{f^\star}$, the end-to-end jitter of $f^\star$ is defined as
\begin{equation}
J_{f^\star}
=
\max_{n\in\mathcal N_{f^\star}}
d^{\mathrm{e2e}}_{f^\star,n}
-
\min_{n\in\mathcal N_{f^\star}}
d^{\mathrm{e2e}}_{f^\star,n}.
\label{eq:jitter}
\end{equation}
For a throughput window $[0,W]$ with $0<W\le R$, the throughput of $f^\star$ is defined as
\begin{equation}
\Theta_{f^\star}(W)
=
\frac{\ell}{W}
\sum_{n\in\mathcal N_{f^\star}}
\mathbf 1
\left\{
t^{\mathrm{dst}}_{f^\star,n}\le W
\right\},
\label{eq:throughput}
\end{equation}
where $\mathbf 1\{\cdot\}$ denotes the indicator function, equal to one when the condition holds and zero otherwise.

The backlog $Q_h(t)$ at output port $p_h$ is the number of packets that have arrived but have not completed serialization by time $t$, given by
\begin{equation}
Q_h(t)
=
\sum_{f\in\mathcal F_h}
\sum_{n\ge0}
\mathbf 1
\left\{
a_{h,f,n}\le t<C_{h,f,n}
\right\}.
\label{eq:port-backlog}
\end{equation}
We seek deterministic upper bounds $D_{f^\star}^{\mathrm U}$, $J_{f^\star}^{\mathrm U}$, and $B_h^{\mathrm U}$ on end-to-end delay, jitter, and per-port backlog, respectively, together with a lower bound $\Theta_{f^\star}^{\mathrm L}(W)$ on throughput.

\subsection{Network Calculus Preliminaries}
\label{subsec:network-calculus-preliminaries}

\begin{figure}[t]
    \centering
    \includegraphics[width=\columnwidth]{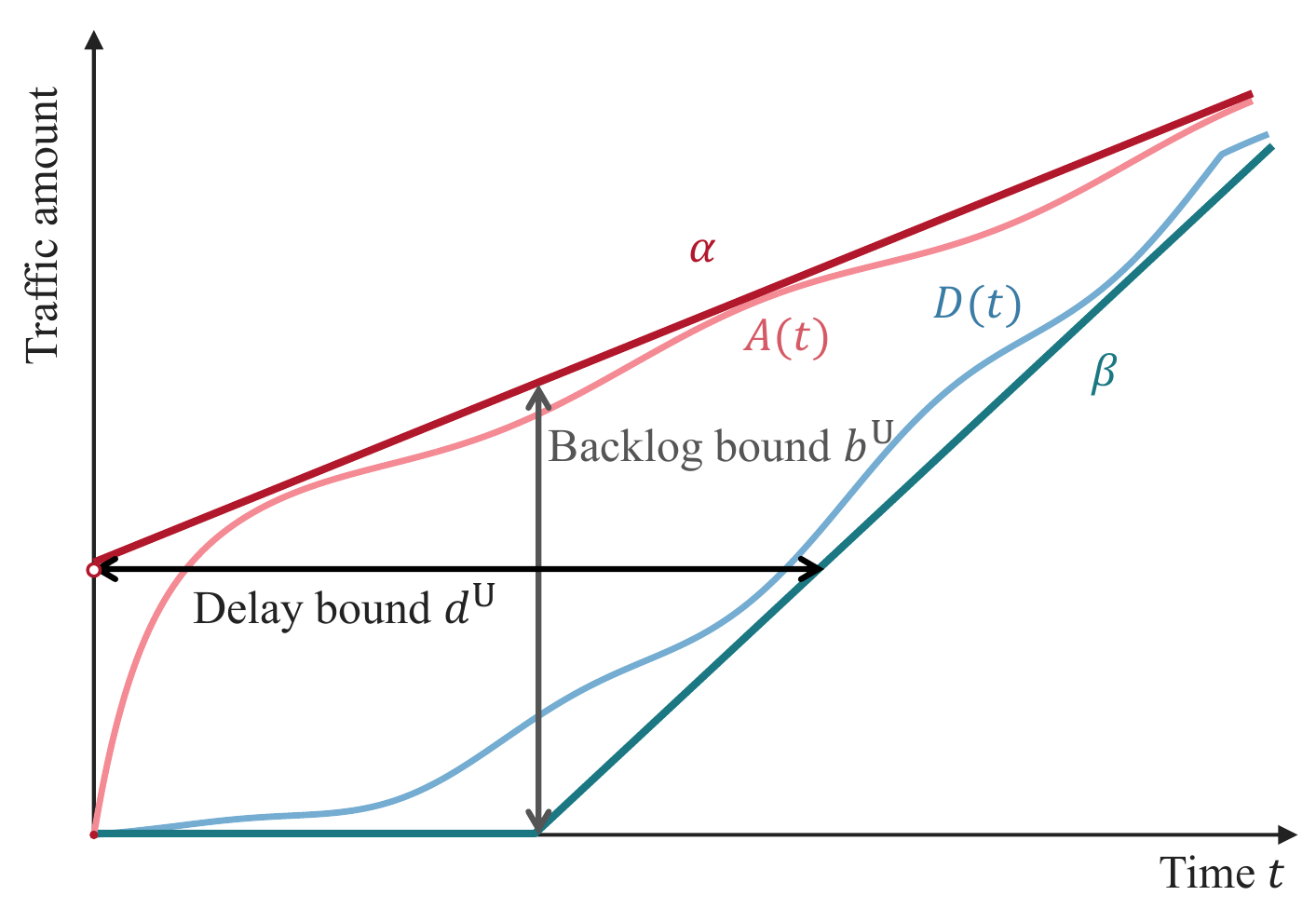}
    \caption{Illustration of delay and backlog bounds in network calculus.}
    \label{fig:nc-preliminaries}
\end{figure}

In deterministic network calculus, traffic entering and leaving a network element is described by the cumulative arrival function $A(t)$ and cumulative departure function $D(t)$, respectively~\cite{leboudec2001networkcalculus}. An arrival bound $\alpha(x)$, conventionally referred to as an arrival curve, limits the amount of traffic arriving within any interval of duration $x$, satisfying
\begin{equation}
A(v)-A(u)
\le
\alpha(v-u),
\qquad 0\le u\le v.
\label{eq:nc-arrival-curve}
\end{equation}
The service guarantee of a network element is characterized by a service curve $\beta(x)$, which bounds cumulative departures from below through
\begin{equation}
\begin{aligned}
D(v)
&\ge (A\otimes\beta)(v)\\
&=
\inf_{0\le u\le v}
\left\{
A(u)+\beta(v-u)
\right\},
\qquad v\ge0,
\end{aligned}
\label{eq:nc-service-curve}
\end{equation}
where $\otimes$ denotes the min-plus convolution.

Two fundamental results in deterministic network calculus are the backlog and delay bounds, which are obtained by combining the arrival bound $\alpha(x)$ and the service guarantee $\beta(x)$~\cite{leboudec2001networkcalculus}, as illustrated in Fig.~\ref{fig:nc-preliminaries}. The backlog upper bound is given by the maximum vertical deviation between $\alpha$ and $\beta$:
\begin{equation}
b^{\mathrm U}
=
\sup_{x\ge0}
\left[
\alpha(x)-\beta(x)
\right]^+,
\label{eq:nc-backlog-bound}
\end{equation}
where $[y]^+=\max\{y,0\}$. Under FIFO service, the delay upper bound is given by their maximum horizontal deviation:
\begin{equation}
d^{\mathrm U}
=
\sup_{x\ge0}
\inf
\left\{
d\ge0:
\alpha(x)\le\beta(x+d)
\right\}.
\label{eq:nc-delay-bound}
\end{equation}

The above formulation characterizes traffic arrivals and service guarantees solely in terms of interval duration. Time-varying network calculus generalizes the arrival bound $\alpha(x)$ and the service guarantee $\beta(x)$ to bivariate functions $\alpha(u,v)$ and $\beta(u,v)$, respectively, where $u$ and $v$ denote the start and end times of an interval with $0\le u\le v$~\cite{daniel2013timevarying,falk2019intermittence}. The corresponding arrival and service constraints are
\begin{align}
A(v)-A(u)
&\le \alpha(u,v),
\label{eq:nc-time-varying-arrival-bound}\\
D(v)
&\ge
\inf_{0\le u\le v}
\left\{
A(u)+\beta(u,v)
\right\}.
\label{eq:nc-time-varying-service}
\end{align}
These bivariate functions preserve the temporal positions of the intervals when traffic arrivals or service guarantees vary over time.

When multiple flows share a network element, their individual arrival bounds can be summed to obtain an aggregate arrival bound, which is combined with the service guarantee to derive local performance bounds. Since traffic characteristics may change after service, per-flow output bounds are derived to characterize the departing traffic. In multihop analysis, these output bounds are used to establish arrival bounds at subsequent network elements, enabling performance analysis along routing paths.

\section{Traffic and Service Analysis for TAS Ports}
\label{sec:tas-port-analysis}

This section conducts a network-calculus analysis for TAS output ports. Specifically, since transmission opportunities depend on the timing of GCL transmission windows, we characterize packet arrivals and GCL service over absolute time intervals. Based on these characterizations, we derive single-port performance bounds and per-flow output bounds for downstream arrival analysis.

\subsection{Absolute-Time Arrival Bounds}
\label{subsec:absolute-time-arrival-bounds}

Under the TAS mechanism, packet waiting times depend on how traffic arrivals align with periodic GCL transmission windows. Arrival bounds based solely on interval duration may be combined with GCL service conditions that cannot occur at the same time, leading to conservative performance bounds. Following the time-varying network calculus formulation introduced in Section~\ref{subsec:network-calculus-preliminaries}, we characterize packet arrivals using an absolute-time arrival bound that depends on the interval start time $u$ and duration $x$.

At output port $p_h$, let $A_{f,h}(t)$ denote the cumulative number of packets of flow $f\in\mathcal F_h$ that have arrived by time $t$. We adopt the right-continuous convention, so arrivals at time $t$ are included in $A_{f,h}(t)$. Since the system is initially empty, $A_{f,h}(0^-)=0$. For any $0\le u\le u+x\le E$, the absolute-time arrival bound $\Gamma_{f,h}^{\mathrm{in}}(u,x)$ satisfies
\begin{equation}
A_{f,h}(u+x)-A_{f,h}(u^-)
\le
\Gamma_{f,h}^{\mathrm{in}}(u,x).
\label{eq:absolute-time-arrival-bound}
\end{equation}
Since $A_{f,h}(u^-)$ excludes arrivals at time $u$, the difference counts arrivals at both endpoints of $[u,u+x]$, including when $x=0$. Summing the arrival bounds of all flows in $\mathcal F_h$ gives an upper bound on the number of packets arriving at output port $p_h$ during $[u,u+x]$:
\begin{equation}
\Gamma_{\Sigma,h}^{\mathrm{in}}(u,x)
=
\sum_{f\in\mathcal F_h}
\Gamma_{f,h}^{\mathrm{in}}(u,x).
\label{eq:aggregate-arrival-bound}
\end{equation}

For each flow $f\in\mathcal F$, we determine its arrival bound at the first output port $p_{s_f}$ it traverses on $\mathcal P_{f^\star}$. If the flow enters directly from its source, its arrival times equal the source release times in \eqref{eq:source-release}. The number of arrivals in $[u,u+x]$ is therefore given by
\begin{equation}
\Gamma_{f,s_f}^{\mathrm{in}}(u,x)
=
\#\left\{
n\in\mathbb Z_{\ge0}:
u
\le
\theta_f+nP_f
\le
u+x
\right\}.
\label{eq:periodic-entry-arrival-bound}
\end{equation}
If the flow enters through upstream output ports and ISLs, its arrival times need not be periodic. In this case, the arrival bound must account for upstream TAS service, ISL propagation, and traffic competing at those output ports.

\subsection{GCL Service Characterization}
\label{subsec:gcl-service-characterization}

To complement the absolute-time arrival bounds, we characterize the service provided under the GCL at output port $p_h$ over an absolute time interval $[u,v]$. From \eqref{eq:tas-start-condition}, the set of feasible serialization start times at $p_h$ is
\begin{equation}
\mathcal E_h
=
\left\{
s:
[s,s+q_h)\subseteq\mathcal O_h
\right\},
\label{eq:feasible-serialization-start-set}
\end{equation}
where $q_h$ is the serialization time at $p_h$ and $\mathcal O_h$ is the set of gate-open times specified by its GCL.

To derive the GCL service, we consider a reference queue that remains backlogged from time $u$. Each packet starts serialization at the earliest feasible time after the preceding packet completes serialization. Let $c_{h,m}(u)$ denote the serialization completion time of the $m$-th packet in the reference queue. For the first packet, this time is given by
\begin{equation}
c_{h,1}(u)
=
\inf
\left\{
s+q_h:
s\ge u,\;
s\in\mathcal E_h
\right\},
\label{eq:first-reference-completion}
\end{equation}
and the subsequent completion times are given recursively by
\begin{equation}
c_{h,m+1}(u)
=
\inf
\left\{
s+q_h:
s\ge c_{h,m}(u),\;
s\in\mathcal E_h
\right\},
\ m\ge1.
\label{eq:reference-completion-recursion}
\end{equation}
If the corresponding feasible set is empty, the completion time is set to $+\infty$. We define the GCL service $S_h(u,v)$ as the number of packets that complete serialization in this reference queue during $[u,v]$:
\begin{equation}
S_h(u,v)
=
\#\left\{
m\ge1:
c_{h,m}(u)\le v
\right\},
\
0\le u\le v\le E.
\label{eq:gcl-service-function}
\end{equation}

We next use the GCL service to bound packet departures from the actual queue at $p_h$. Let $D_{f,h}(t)$ denote the cumulative number of packets of flow $f\in\mathcal F_h$ that have completed serialization at $p_h$ by time $t$. Following the right-continuous convention for $A_{f,h}(t)$, departures at time $t$ are included in $D_{f,h}(t)$, and $D_{f,h}(0^-)=0$ under the empty initial state. The cumulative arrivals and departures of all flows at $p_h$ are $A_{\Sigma,h}(t)=\sum_{f\in\mathcal F_h}A_{f,h}(t)$ and $D_{\Sigma,h}(t)=\sum_{f\in\mathcal F_h}D_{f,h}(t)$. The following theorem establishes a lower bound on the cumulative packet departures from the actual queue using the GCL service $S_h(u,v)$.

\begin{theorem}[Lower Bound on Cumulative Packet Departures under GCL Service]
\label{thm:gcl-cumulative-departure-lower-bound}
Under the TAS mechanism described in Section~\ref{subsec:tas-mechanism}, the cumulative packet departures at $p_h$ satisfy
\begin{equation}
D_{\Sigma,h}(v)
\ge
\inf_{0\le u\le v}
\left\{
A_{\Sigma,h}(u^-)+S_h(u,v)
\right\},
\quad
v\in\mathcal J.
\label{eq:gcl-cumulative-departure-lower-bound}
\end{equation}
\end{theorem}

\begin{proof}
If the queue is empty after all events at time $v$, all packets arriving by time $v$ have departed, so $D_{\Sigma,h}(v)=A_{\Sigma,h}(v)\ge A_{\Sigma,h}(v^-)$. Since $S_h(v,v)=0$, choosing $u=v$ establishes the lower bound.

Otherwise, let $u$ be the start time of the busy period containing $v$. The queue is empty immediately before $u$ and remains backlogged throughout $[u,v]$. Hence, all packets arriving strictly before $u$ have already departed. Since the queue remains backlogged, packets start serialization at the earliest feasible times, as in the reference queue. Thus, at least $S_h(u,v)$ packets depart during $[u,v]$. Therefore,
\begin{equation}
D_{\Sigma,h}(v)
-
A_{\Sigma,h}(u^-)
\ge
S_h(u,v).
\label{eq:continuous-backlog-service}
\end{equation}
Since this $u$ is included in the infimum in \eqref{eq:gcl-cumulative-departure-lower-bound}, the result follows.
\end{proof}

\subsection{Delay and Backlog Bounds at a TAS Output Port}
\label{subsec:delay-backlog-bounds}

\begin{figure}[t]
    \centering
    \includegraphics[width=\columnwidth]{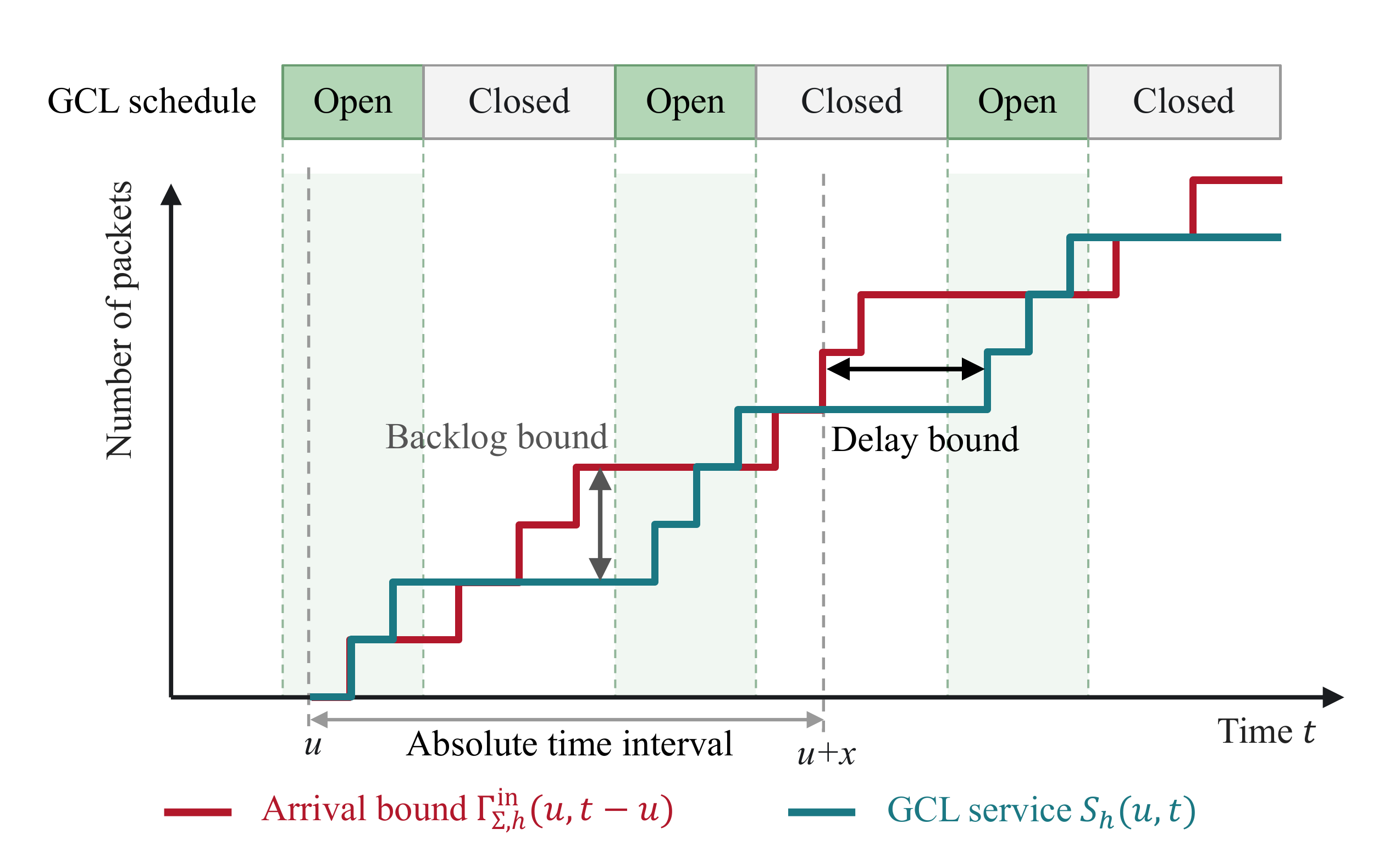}
    \caption{Illustration of absolute-time network calculus at a TAS output port.}
    \label{fig:absolute-time-analysis}
\end{figure}

Using the arrival bounds and GCL service established above, we derive bounds on delay and backlog at output port $p_h$, as illustrated in Fig.~\ref{fig:absolute-time-analysis}.

\begin{theorem}[Single-Port Delay and Backlog Bounds]
\label{thm:single-port-delay-backlog}
For packets arriving at $p_h$ at time $t\in\mathcal J$, the delay until completion of serialization is bounded by
\begin{align}
d_h^{\mathrm U}(t)
=
\inf\Bigl\{
d\in[0,E-t]:
&\;
\Gamma_{\Sigma,h}^{\mathrm{in}}(u,t-u)
\nonumber\\
&\hspace{-2em}\le
S_h(u,t+d),\;
\forall u\in[0,t]
\Bigr\}.
\label{eq:port-delay-bound}
\end{align}
If the feasible set is empty, we set $d_h^{\mathrm U}(t)=+\infty$. When $d_h^{\mathrm U}(t)<+\infty$, all packets arriving no later than $t$ complete serialization by $t+d_h^{\mathrm U}(t)$.

The backlog at $p_h$ at time $t$ satisfies
\begin{equation}
Q_h(t)
\le
\sup_{0\le u\le t}
\left[
\Gamma_{\Sigma,h}^{\mathrm{in}}(u,t-u)
-
S_h(u,t)
\right]^+,
\label{eq:port-backlog-bound}
\end{equation}
where $[x]^+=\max\{x,0\}$.
\end{theorem}

\begin{proof}
For any feasible $d$ in \eqref{eq:port-delay-bound}, the arrival bounds give $A_{\Sigma,h}(u^-)+S_h(u,t+d)\ge A_{\Sigma,h}(t)$ for every $u\in[0,t]$. This inequality also holds for $u\in(t,t+d]$, since $A_{\Sigma,h}(u^-)\ge A_{\Sigma,h}(t)$ and $S_h(u,t+d)\ge0$. Applying \eqref{eq:gcl-cumulative-departure-lower-bound} gives $D_{\Sigma,h}(t+d)\ge A_{\Sigma,h}(t)$. Under FIFO service, all packets arriving no later than $t$ therefore complete serialization by $t+d$. Taking the infimum over feasible $d$ and using the right-continuity of the departure process proves the delay bound.

For backlog, $Q_h(t)=A_{\Sigma,h}(t)-D_{\Sigma,h}(t)$. Combining \eqref{eq:gcl-cumulative-departure-lower-bound} with the arrival bounds in \eqref{eq:absolute-time-arrival-bound} directly gives \eqref{eq:port-backlog-bound}.
\end{proof}

\subsection{Per-Flow Output Bounds}
\label{subsec:per-flow-output-bounds}

For a flow traversing both $p_h$ and $p_{h+1}$, its packets leave $p_h$ upon completing serialization and arrive at $p_{h+1}$ after propagation over ISL $e_h$. To bound the number of packets leaving $p_h$ during a given interval, we need to determine when these packets could have arrived at $p_h$. We therefore derive a delay bound for packets completing serialization within $\mathcal J$ and use it to establish a per-flow output bound.

The number of packets arriving at $p_h$ from all flows in $\mathcal F_h$ during any interval of length $x$ within $\mathcal J$ is bounded by
\begin{equation}
\alpha_{\Sigma,h}(x)
=
\sup_{0\le u\le E-x}
\Gamma_{\Sigma,h}^{\mathrm{in}}(u,x),
\quad
0\le x\le E.
\label{eq:duration-based-aggregate-arrival-bound}
\end{equation}
We assume $\alpha_{\Sigma,h}(x)<+\infty$ for all $x\in[0,E]$. For the reference queue used to characterize the GCL service, $c_{h,m}(u)-u$ is the time required to complete serialization of $m$ packets when the queue remains backlogged from time $u$. Since the GCL is periodic, taking the supremum over one period covers all possible start times within a GCL cycle:
\begin{equation}
\sigma_h(m)
=
\sup_{u\in[0,T)}
\left[
c_{h,m}(u)-u
\right],
\quad
m\in\mathbb Z_{>0},
\label{eq:phase-independent-reference-completion-bound}
\end{equation}
with $\sigma_h(0)=0$. Using these two bounds, we derive the following delay bound.

\begin{proposition}[Delay Bound for Packets Completing Serialization at a TAS Output Port]
\label{prop:completed-packet-delay-bound}
For any packet arriving at $p_h$ at time $a\in\mathcal J$ and completing serialization at time $c\le E$,
\begin{equation}
q_h
\le
c-a
\le
\bar d_h,
\label{eq:completed-packet-delay-bounds}
\end{equation}
where
\begin{equation}
\bar d_h
=
\max
\left\{
0,\;
\sup_{0\le x\le E}
\left[
\sigma_h\!\left(
\left\lfloor
\alpha_{\Sigma,h}(x)
\right\rfloor
\right)
-
x
\right]
\right\}.
\label{eq:uniform-packet-delay-bound}
\end{equation}
Unlike $d_h^{\mathrm U}(t)$ in Section~\ref{subsec:delay-backlog-bounds}, which depends on the arrival time $t$, $\bar d_h$ applies to all packets completing serialization within $\mathcal J$. In addition, a packet arriving at time $a\in\mathcal J$ completes serialization by $a+\bar d_h$ if $a+\bar d_h\le E$.
\end{proposition}

\begin{proof}
Consider a packet arriving at $p_h$ at time $a\in\mathcal J$. To account for packets that may complete serialization after time $E$, we consider an extended system in which the GCL continues periodically and no new packets arrive after $E$. Let $u$ be the start time of the busy period containing $a$, and set $x=a-u$. All packets served before the target packet in this busy period, together with the target packet itself, must have arrived during $[u,a]$. Their total number is therefore at most $\lfloor\alpha_{\Sigma,h}(x)\rfloor$.

In the extended system, the queue remains backlogged from $u$ until the target packet finishes serialization. By \eqref{eq:phase-independent-reference-completion-bound} and \eqref{eq:uniform-packet-delay-bound}, its serialization completion time $\widetilde c$ in the extended system satisfies
\begin{align}
\widetilde c
&\le
u+\sigma_h\!\left(
\lfloor\alpha_{\Sigma,h}(x)\rfloor
\right)
\nonumber\\
&\le u+x+\bar d_h = a+\bar d_h.
\end{align}
Since the original and extended systems behave identically over $\mathcal J$, any packet completing serialization at time $c\le E$ in the original system satisfies $c=\widetilde c\le a+\bar d_h$. Conversely, if $a+\bar d_h\le E$, then $\widetilde c\le E$, which guarantees that the packet also completes serialization within $\mathcal J$ at $c=\widetilde c$. Finally, the lower bound $c-a\ge q_h$ holds because serializing a packet takes time $q_h$.
\end{proof}

A packet completing serialization during $[s,t]$ must have arrived no earlier than $s-\bar d_h$ and no later than $t-q_h$. Applying the arrival bound of each flow to this interval gives the following result.

\begin{theorem}[Per-Flow Output Bound at a TAS Output Port]
\label{thm:per-flow-output-bound}
For any flow $f\in\mathcal F_h$ and closed interval $[s,t]\subseteq\mathcal J$, the cumulative departures satisfy
\begin{equation}
D_{f,h}(t)
-
D_{f,h}(s^-)
\le
\Gamma_{f,h}^{\mathrm{out}}(s,t-s),
\label{eq:per-flow-output-bound-inequality}
\end{equation}
where $l=\max\{0,s-\bar d_h\}$, $r=t-q_h$, and
\begin{equation}
\Gamma_{f,h}^{\mathrm{out}}(s,t-s)
=
\begin{cases}
\Gamma_{f,h}^{\mathrm{in}}(l,r-l),
& l\le r,\\
0,
& l>r.
\end{cases}
\label{eq:per-flow-output-bound}
\end{equation}
\end{theorem}

\begin{proof}
Consider any packet of flow $f$ that completes serialization at time $c\in[s,t]$, and let $a$ denote its arrival time. Since $c\le t\le E$, Proposition~\ref{prop:completed-packet-delay-bound} yields $q_h\le c-a\le\bar d_h$. Combining this inequality with $s\le c\le t$ gives $s-\bar d_h\le a\le t-q_h$. Because no packets arrive before time $0$, we also have $a\ge0$, so the arrival time $a$ must lie in $[l,r]$. If $l>r$, this interval is empty and no packet of flow $f$ can complete serialization during $[s,t]$. Otherwise, every such packet arrives during $[l,r]$, and applying \eqref{eq:absolute-time-arrival-bound} over $[l,r]$ bounds the number of such packets by $\Gamma_{f,h}^{\mathrm{in}}(l,r-l)$, which establishes \eqref{eq:per-flow-output-bound-inequality}.
\end{proof}

\section{Multihop Analysis and End-to-End Performance Bounds}
\label{sec:end-to-end-analysis}

Building on the single-port analysis in Section~\ref{sec:tas-port-analysis}, we extend the analysis to multihop transmission along the routing path. Specifically, we first transform per-flow output bounds into arrival bounds at the next output port by accounting for time-varying ISL propagation. Packet ordering at each output port is then exploited to further tighten the performance bounds. Finally, we derive deterministic bounds on the end-to-end delay, jitter, and throughput of $f^\star$ and the backlog at each output port.

\subsection{Arrival Bound Transformation under Time-Varying ISL Propagation}
\label{subsec:arrival-bound-transformation}

To obtain an arrival bound at $p_{h+1}$ from the per-flow output bound at $p_h$, we account for the time-varying propagation delay of ISL $e_h$. Consider two packets of the same flow that complete serialization at $p_h$ at times $s<t$. By \eqref{eq:propagation-map}, their arrival times at $p_{h+1}$ are $\phi_h(s)$ and $\phi_h(t)$, respectively. The time separation between these arrivals is $\phi_h(t)-\phi_h(s)=(t-s)+\delta_h(t)-\delta_h(s)$. When the propagation delay is constant, this separation equals $t-s$. Otherwise, the arrival interval may become shorter or longer than the corresponding output interval. Directly applying the output bound to the arrival interval may therefore lead to an invalid or unnecessarily loose bound. We instead map each arrival interval at $p_{h+1}$ back to the corresponding interval of serialization completion times at $p_h$.

For an arrival interval $[u,u+x]\subseteq\mathcal J$ at $p_{h+1}$, the corresponding interval of serialization completion times at $p_h$ is
\begin{equation}
\mathcal I_h^{\phi}(u,x)
=
\phi_h^{-1}\!\left(
[u,u+x]\cap\phi_h(\mathcal J)
\right).
\label{eq:isl-inverse-window}
\end{equation}
Since $\phi_h$ is continuous and strictly increasing over $\mathcal J$, its inverse is well defined on $\phi_h(\mathcal J)$, and $\mathcal I_h^{\phi}(u,x)$ is either empty or a closed interval. When nonempty, let $l_h^{\phi}(u,x)$ and $r_h^{\phi}(u,x)$ denote its left and right endpoints, respectively.

\begin{proposition}[Per-Flow Arrival Bound under Time-Varying ISL Propagation]
\label{prop:isl-arrival-transformation}
For a flow $f$ traversing both $p_h$ and $p_{h+1}$, an arrival bound at $p_{h+1}$ is given by
\begin{equation}
\Gamma_{f,h+1}^{\mathrm{in}}(u,x)
=
\begin{cases}
\Gamma_{f,h}^{\mathrm{out}}(l_h^{\phi},r_h^{\phi}-l_h^{\phi}),
& \mathcal I_h^{\phi}(u,x)\ne\varnothing,\\
0,
& \text{otherwise},
\end{cases}
\label{eq:isl-arrival-transformation}
\end{equation}
where $l_h^{\phi}=l_h^{\phi}(u,x)$ and $r_h^{\phi}=r_h^{\phi}(u,x)$.
\end{proposition}

\begin{proof}
Consider any packet of flow $f$ arriving at $p_{h+1}$ during $[u,u+x]$, and let $s$ denote its serialization completion time at $p_h$. Since no packets are in flight at time $0$ and propagation delays are nonnegative, we have $0\le s\le\phi_h(s)\le u+x\le E$. By \eqref{eq:isl-inverse-window}, its serialization completion time must therefore lie in $\mathcal I_h^{\phi}(u,x)$.

If $\mathcal I_h^{\phi}(u,x)$ is empty, no such packet can arrive. Otherwise, every such packet must have completed serialization at $p_h$ during $[l_h^{\phi},r_h^{\phi}]$. Applying \eqref{eq:per-flow-output-bound-inequality} to this interval bounds the number of arriving packets by $\Gamma_{f,h}^{\mathrm{out}}(l_h^{\phi},r_h^{\phi}-l_h^{\phi})$, establishing \eqref{eq:isl-arrival-transformation}.
\end{proof}

The single-port analysis and arrival bound transformation can now be applied successively along the routing path. At each output port, we use the arrival bounds and GCL service to derive delay, backlog, and per-flow output bounds. The output bounds are then transformed into arrival bounds at the next output port using \eqref{eq:isl-arrival-transformation}. For flows entering $\mathcal P_{f^\star}$ through upstream output ports and ISLs, the same analysis is applied along their upstream paths, including competing traffic at those ports, to obtain their arrival bounds at $p_{s_f}$.

\subsection{Tightening Performance Bounds Using Packet Ordering}
\label{subsec:packet-ordering-refinement}

At a TAS output port, a packet must wait for any unfinished packets ahead of it in the FIFO queue. We use packet timing bounds to exclude packets that cannot delay the target packet. The remaining packet count and the single-port backlog bound each limit the number of packets that need to be served. We take the smaller bound and use the GCL service to bound the serialization completion time of the target packet.

For each flow $f\in\mathcal F_h$, let $\mathcal M_f=\{j\in\mathbb Z_{\ge0}:r_{f,j}^{\mathrm{src}}\in\mathcal J\}$ denote the indices of packets released during $\mathcal J$. For each packet $(f,n)$ with $n\in\mathcal M_f$, let $a_{h,f,n}^{\mathrm L}$ denote a lower bound on its arrival time at $p_h$, satisfying $r_{f,n}^{\mathrm{src}}\le a_{h,f,n}^{\mathrm L}\le a_{h,f,n}$. For packets guaranteed to arrive within $\mathcal J$, we also obtain an upper bound $a_{h,f,n}^{\mathrm U}$ such that $a_{h,f,n}\in[a_{h,f,n}^{\mathrm L},a_{h,f,n}^{\mathrm U}]\subseteq\mathcal J$. If flow $f$ enters $p_{s_f}$ directly from its source, then $a_{s_f,f,n}^{\mathrm L}=a_{s_f,f,n}^{\mathrm U}=r_{f,n}^{\mathrm{src}}$; otherwise, these bounds are obtained from its upstream path.

We first exclude packets that cannot precede $(f,n)$ in FIFO order. Since FIFO service and ISL propagation preserve the source order of packets belonging to the same flow, $a_{h,f,i}\le a_{h,f,j}$ and $C_{h,f,i}\le C_{h,f,j}$ for $i<j$. Thus, any packet $(f,j)$ with $j>n$ cannot precede $(f,n)$. Similarly, a packet $(g,j)$ satisfying $a_{h,g,j}^{\mathrm L}>a_{h,f,n}^{\mathrm U}$ must arrive later and cannot delay $(f,n)$.

We also exclude packets that have completed serialization at $p_h$ before $(f,n)$ arrives, using upper bounds on their serialization completion times at this port. From the feasible serialization start times $\mathcal E_h$ in \eqref{eq:feasible-serialization-start-set}, the set of feasible serialization completion times is $\mathcal T_h^{\mathrm C}=\{s+q_h:s\in\mathcal E_h\}$.
For a packet $(f,n)$ arriving within $[a_{h,f,n}^{\mathrm L},a_{h,f,n}^{\mathrm U}]$, serialization cannot finish before $a_{h,f,n}^{\mathrm L}+q_h$. Theorem~\ref{thm:single-port-delay-backlog} and Proposition~\ref{prop:completed-packet-delay-bound} provide the upper bounds $a_{h,f,n}^{\mathrm U}+d_h^{\mathrm U}(a_{h,f,n}^{\mathrm U})$ and $a_{h,f,n}^{\mathrm U}+\bar d_h$, respectively, for completions within $\mathcal J$. Combining these bounds with the feasible GCL completion times gives an initial interval of possible serialization completion times for packet $(f,n)$ within $\mathcal J$:
\begin{equation}
\begin{aligned}
\mathcal C_{h,f,n}^{\mathrm I}
=
\operatorname{hull}\Bigl(
\mathcal T_h^{\mathrm C}
\cap
\bigl[
&a_{h,f,n}^{\mathrm L}+q_h,\\
&\min\bigl\{
E,\,
a_{h,f,n}^{\mathrm U}+\bar d_h,\\
&\qquad
a_{h,f,n}^{\mathrm U}
+d_h^{\mathrm U}(a_{h,f,n}^{\mathrm U})
\bigr\}
\bigr]
\Bigr).
\end{aligned}
\label{eq:initial-gcl-completion-interval}
\end{equation}
Here, $\operatorname{hull}(\cdot)$ denotes the smallest closed interval containing the given set, with an empty set remaining empty.

To exclude a packet based on its completion time, we must guarantee that it completes serialization within $\mathcal J$. Let $\mathcal M_{h,g}^{\mathrm C}\subseteq\mathcal M_g$ contain the indices $i$ of packets whose arrival at $p_h$ within $\mathcal J$ is established and for which either $d_h^{\mathrm U}(a_{h,g,i}^{\mathrm U})<+\infty$ or $a_{h,g,i}^{\mathrm U}+\bar d_h\le E$. These conditions guarantee completion within $\mathcal J$, so $\max\mathcal C_{h,g,i}^{\mathrm I}$ is a valid upper bound on the serialization completion time of each such packet.

Since packets of the same flow complete serialization in source order, we define an upper bound $\widehat C_{h,g,j}^{\mathrm U}$ on the serialization completion time of packet $(g,j)$ using the completion-time bounds of that packet and its successors:
\begin{equation}
\widehat C_{h,g,j}^{\mathrm U}
=
\min_{\substack{i\in\mathcal M_{h,g}^{\mathrm C}\\i\ge j}}
\max\mathcal C_{h,g,i}^{\mathrm I},
\label{eq:certified-prior-completion}
\end{equation}
where the minimum is $+\infty$ if the index set is empty. If $\widehat C_{h,g,j}^{\mathrm U}<a_{h,f,n}^{\mathrm L}$, packet $(g,j)$ must have completed serialization before $(f,n)$ arrives.

Combining the FIFO ordering and serialization completion-time conditions, we define the set $\mathcal X_{h,f,n}$ of packets that can be excluded when analyzing packet $(f,n)$ at output port $p_h$:
\begin{equation}
\begin{aligned}
\mathcal X_{h,f,n}
=\bigl\{(g,j):\;
&g\in\mathcal F_h,\;
j\in\mathcal M_g,\\
&(g=f\ \text{and}\ j>n)\\
&\quad\text{or}\quad
a_{h,g,j}^{\mathrm L}>a_{h,f,n}^{\mathrm U}\\
&\quad\text{or}\quad
\widehat C_{h,g,j}^{\mathrm U}<a_{h,f,n}^{\mathrm L}
\bigr\}.
\end{aligned}
\label{eq:packet-exclusion-set}
\end{equation}

From \eqref{eq:port-backlog-bound}, the backlog at $p_h$ over $\mathcal J$ is bounded by
\begin{equation}
\overline B_h
=
\sup_{0\le u\le t\le E}
\left[
\Gamma_{\Sigma,h}^{\mathrm{in}}(u,t-u)
-
S_h(u,t)
\right]^+.
\label{eq:single-port-backlog-upper-bound}
\end{equation}
When packet $(f,n)$ arrives, only the packet itself and its unfinished FIFO predecessors need to be served. These packets contribute to the backlog at $p_h$ and lie outside $\mathcal X_{h,f,n}$, so their number is bounded by both the single-port backlog bound and the number of packets remaining after exclusion. Taking the smaller bound gives
\begin{equation}
k_{h,f,n}
=
\min\left\{
\left\lfloor\overline B_h\right\rfloor,\,
\sum_{g\in\mathcal F_h}|\mathcal M_g|
-
|\mathcal X_{h,f,n}|
\right\}.
\label{eq:fifo-prefix-workload}
\end{equation}
Here, $k_{h,f,n}$ includes packet $(f,n)$ itself.

The completion times $c_{h,m}(u)$ of the reference queue, defined in Section~\ref{subsec:gcl-service-characterization}, allow us to bound the time needed to serve these packets. Since $(f,n)$ arrives no later than $a_{h,f,n}^{\mathrm U}$ and at most $k_{h,f,n}$ packets need to be served through its completion, $c_{h,k_{h,f,n}}(a_{h,f,n}^{\mathrm U})$ provides an additional upper bound on its serialization completion time within $\mathcal J$. Combining this bound with the initial interval gives a tightened interval of possible serialization completion times for packet $(f,n)$:
\begin{equation}
\begin{aligned}
\mathcal C_{h,f,n}^{\mathrm T}
=
\operatorname{hull}\Bigl(
&\mathcal C_{h,f,n}^{\mathrm I}
\cap\mathcal T_h^{\mathrm C}\\
&\cap
(-\infty,
c_{h,k_{h,f,n}}(a_{h,f,n}^{\mathrm U})]
\Bigr).
\end{aligned}
\label{eq:local-refined-completion-interval}
\end{equation}
The refined interval spans the earliest and latest GCL-feasible completion times within the initial interval that satisfy the packet-ordering bound.

\begin{theorem}[Packet Timing Bounds Using Packet Ordering]
\label{thm:packet-ordering-refinement}
For packet $(f,n)$ arriving at $p_h$ within $[a_{h,f,n}^{\mathrm L},a_{h,f,n}^{\mathrm U}]\subseteq\mathcal J$, if it completes serialization within $\mathcal J$, then
\begin{equation}
C_{h,f,n}
\in
\mathcal C_{h,f,n}^{\mathrm T}.
\label{eq:packet-ordering-completion-interval-guarantee}
\end{equation}
Furthermore, if $c_{h,k_{h,f,n}}(a_{h,f,n}^{\mathrm U})\le E$, its completion within $\mathcal J$ is guaranteed, with
\begin{equation}
C_{h,f,n}
\le
c_{h,k_{h,f,n}}(a_{h,f,n}^{\mathrm U}).
\label{eq:refined-serialization-bound}
\end{equation}
\end{theorem}

\begin{proof}
Let $a\in[a_{h,f,n}^{\mathrm L},a_{h,f,n}^{\mathrm U}]$ be the actual arrival time of packet $(f,n)$, and let $M(a)$ denote the number of unfinished packets that precede or equal $(f,n)$ in FIFO order, including the packet itself. These packets contribute to the backlog at $p_h$ and lie outside $\mathcal X_{h,f,n}$, so $1\le M(a)\le k_{h,f,n}$.

At time $a$, the remaining work before $(f,n)$ completes requires no more than $M(a)$ full packet serializations. Under FIFO service, later arrivals cannot overtake $(f,n)$. Since the completion times $c_{h,m}(u)$ of the reference queue are nondecreasing in both the packet count and the starting time,
\begin{equation}
c_{h,M(a)}(a)
\le
c_{h,k_{h,f,n}}(a_{h,f,n}^{\mathrm U}).
\end{equation}
If $c_{h,k_{h,f,n}}(a_{h,f,n}^{\mathrm U})\le E$, comparison with the GCL reference queue guarantees completion within $\mathcal J$ and proves \eqref{eq:refined-serialization-bound}. Otherwise, any completion within $\mathcal J$ satisfies $C_{h,f,n}\le E<c_{h,k_{h,f,n}}(a_{h,f,n}^{\mathrm U})$. Thus, every completion within $\mathcal J$ satisfies the additional upper bound $c_{h,k_{h,f,n}}(a_{h,f,n}^{\mathrm U})$. Combining this bound with $\mathcal C_{h,f,n}^{\mathrm I}$ and $\mathcal T_h^{\mathrm C}$ proves \eqref{eq:packet-ordering-completion-interval-guarantee}.
\end{proof}

Let $C_{h,f,n}^{\mathrm L}$ and $C_{h,f,n}^{\mathrm U}$ denote the endpoints of the nonempty interval $\mathcal C_{h,f,n}^{\mathrm T}$. For a packet $(f,n)$ traversing both $p_h$ and $p_{h+1}$, if it is guaranteed to complete serialization at $p_h$ within $\mathcal J$ and $\phi_h(C_{h,f,n}^{\mathrm U})\le E$, its arrival-time bounds at $p_{h+1}$ are $a_{h+1,f,n}^{\mathrm L}=\phi_h(C_{h,f,n}^{\mathrm L})$ and $a_{h+1,f,n}^{\mathrm U}=\phi_h(C_{h,f,n}^{\mathrm U})$. Applying this refinement at successive output ports provides refined arrival and completion-time bounds along the routing path, which are then used to tighten the end-to-end performance bounds.

\subsection{Deterministic End-to-End Performance Bounds}
\label{subsec:end-to-end-performance-bounds}

Combining the single-port analysis, the arrival bound transformation under time-varying ISL propagation, and the performance bounds tightened using packet ordering, we derive deterministic bounds on the end-to-end delay, jitter, and throughput of the flow of interest $f^\star$ and the backlog at each output port.

\begin{theorem}[Deterministic Performance Bounds]
\label{thm:deterministic-performance-bounds}
If $\mathcal N_{f^\star}$ is nonempty and every target packet $n\in\mathcal N_{f^\star}$ is guaranteed to reach the destination within $\mathcal J$, the end-to-end delay of $f^\star$ is bounded by
\begin{equation}
D_{f^\star}^{\mathrm U}
=
\max_{n\in\mathcal N_{f^\star}}
\left\{
\phi_H(C_{H,f^\star,n}^{\mathrm U})
-
r_{f^\star,n}^{\mathrm{src}}
\right\}.
\label{eq:final-end-to-end-delay-bound}
\end{equation}
The corresponding end-to-end jitter bound is
\begin{equation}
J_{f^\star}^{\mathrm U}
=
D_{f^\star}^{\mathrm U}
-
\min_{n\in\mathcal N_{f^\star}}
\left\{
\phi_H(C_{H,f^\star,n}^{\mathrm L})
-
r_{f^\star,n}^{\mathrm{src}}
\right\}.
\label{eq:final-jitter-bound}
\end{equation}
The throughput over $[0,W]$, where $0<W\le R$, is bounded from below by
\begin{equation}
\Theta_{f^\star}^{\mathrm L}(W)
=
\frac{\ell}{W}
\sum_{n\in\mathcal N_{f^\star}}
\mathbf 1
\left\{
\phi_H(C_{H,f^\star,n}^{\mathrm U})\le W
\right\}.
\label{eq:final-throughput-bound}
\end{equation}

For each packet $(f,n)$, set $C_{h,f,n}^{\mathrm U,*}=C_{h,f,n}^{\mathrm U}$ if a bound $C_{h,f,n}\le C_{h,f,n}^{\mathrm U}\le E$ has been obtained, and $C_{h,f,n}^{\mathrm U,*}=+\infty$ otherwise. The following per-port backlog bound combines the single-port backlog bound $\overline B_h$ with the packet-level timing bounds and holds independently of the conditions for the end-to-end bounds:
\begin{equation}
\begin{gathered}
\sup_{t\in\mathcal J}Q_h(t)\le B_h^{\mathrm U},\\
B_h^{\mathrm U}
=\min\!\left\{
\overline B_h,\,
\sup_{t\in\mathcal J}
\sum_{\substack{f\in\mathcal F_h\\n\in\mathcal M_f}}
\mathbf 1\{a_{h,f,n}^{\mathrm L}\le t<C_{h,f,n}^{\mathrm U,*}\}
\right\}.
\end{gathered}
\label{eq:final-backlog-bound}
\end{equation}
\end{theorem}

\begin{proof}
For each target packet, its serialization completion-time bounds at $p_H$ and the monotonicity of $\phi_H$ give lower and upper bounds on its reception time at the destination. Since $a_{1,f^\star,n}=r_{f^\star,n}^{\mathrm{src}}$, the delay and jitter bounds follow directly from \eqref{eq:end-to-end-delay} and \eqref{eq:jitter}. The throughput bound follows from \eqref{eq:throughput} by counting target packets whose upper reception-time bounds do not exceed $W$.

For backlog, every packet counted by $Q_h(t)$ was released during $\mathcal J$ and satisfies $a_{h,f,n}^{\mathrm L}\le t<C_{h,f,n}^{\mathrm U,*}$. Thus, the indicator sum in \eqref{eq:final-backlog-bound} provides an upper bound on $Q_h(t)$. Taking the supremum over $\mathcal J$ and combining it with the single-port backlog bound $\overline B_h$ gives \eqref{eq:final-backlog-bound}.
\end{proof}

\subsection{Performance Bound Computation and Complexity Analysis}
\label{subsec:implementation-complexity}

The performance bounds derived above involve extrema over continuous time intervals and successive transformations of traffic bounds along the routing path. This subsection describes how these bounds are computed while preserving their validity and analyzes the associated computational complexity. The computation proceeds along the routing path, with the upstream output ports of cross flows analyzed before their downstream ports.

At each output port, we partition the evaluation interval $\mathcal{J}$ into subintervals to compute the single-port delay bound in \eqref{eq:port-delay-bound} and the backlog bound in \eqref{eq:single-port-backlog-upper-bound}. The endpoints of these subintervals are used to enlarge arrival intervals and to obtain conservative GCL service estimates by considering later service start times for the delay bound and shorter service intervals for the backlog bound. By the monotonicity of the arrival bounds and GCL service with respect to interval inclusion, this procedure accounts for all possible values of the time variables within the subintervals. For the uniform packet delay bound in \eqref{eq:uniform-packet-delay-bound}, we construct an upper bound on $\alpha_{\Sigma,h}(x)$ using the flow periods and packet arrival-time bounds and examine the interval lengths at which this upper bound changes. To compute $\sigma_h(m)$, we evaluate the serialization completion times of the reference queue at critical start times within one GCL period, as determined by the GCL windows and packet serialization time, and consider the corresponding right-hand limits.

For the arrival bound transformation in \eqref{eq:isl-arrival-transformation}, we obtain the possible arrival interval of each packet by mapping its departure interval through $\phi_h$ in \eqref{eq:propagation-map}. Since $\phi_h$ is continuous and strictly increasing, the mapped interval intersects a given arrival interval if and only if the departure interval intersects the corresponding inverse image; hence, packet arrivals can be bounded through forward mapping without explicitly inverting $\phi_h$. For packet-ordering refinement, the ordering of arrival and completion-time bounds within each flow enables the exclusion conditions in \eqref{eq:packet-exclusion-set} to be evaluated via binary search.

To reduce the computational cost, we bound aggregate arrivals from above and GCL service from below using linear functions of the interval length. When the long-term guaranteed service rate exceeds the aggregate arrival rate, the service lower bound exceeds the arrival upper bound beyond a finite interval length. For any interval longer than this threshold, the backlog term in \eqref{eq:single-port-backlog-upper-bound} is zero and the delay condition in \eqref{eq:port-delay-bound} is automatically satisfied for all nonnegative delays, so all such intervals can be excluded from the search. For a fixed number of GCL windows per period and a fixed maximum number of packets that can be transmitted per window, the worst-case time complexity is
\begin{equation}
O\!\left(H_{\mathrm{tot}}F_{\max}(K^2+NK)\log(N+1)\right),
\end{equation}
where $H_{\mathrm{tot}}$ is the total number of analyzed output ports, and $F_{\max}$, $N$, and $K$ are the maximum numbers of flows, packets, and subintervals per output port over $\mathcal{J}$, respectively.

\section{Numerical Results}

This section evaluates the proposed network-calculus method under different network and traffic configurations. We first present the simulation setup and compare the resulting deterministic performance bounds with those of three baselines and with packet-level simulation results. The impact of time-varying ISL propagation on downstream arrival bounds and the effectiveness of packet ordering in tightening performance bounds are then examined. Finally, we conduct a sensitivity analysis to assess how GCL window length and cross-traffic load affect performance bounds.

\subsection{Simulation Setup}
\label{subsec:simulation-setup}

\begin{table*}[t]
\caption{Simulation Parameters}
\label{tab:simulation-parameters}
\centering
\footnotesize
\setlength{\tabcolsep}{4pt}
\begin{tabular}{@{}clc@{\hspace{1em}}clc@{}}
\toprule
\textbf{Symbol} & \textbf{Meaning} & \textbf{Value}
& \textbf{Symbol} & \textbf{Meaning} & \textbf{Value} \\
\midrule
$N_P$ & Number of orbital planes & $72$
& $C_h$ & Link rate & $1~\mathrm{Gbps}$ \\
$N_S$ & Satellites per plane & $22$
& $T$ & GCL period & $8~\mathrm{ms}$ \\
$h_{\mathrm{orb}}$ & Orbital altitude & $550~\mathrm{km}$
& $M_h$ & Number of GCL windows & $2$ \\
$\iota$ & Orbital inclination & $53^\circ$
& $[a_{h,1},b_{h,1})$ & GCL window 1 & $[0,0.036)~\mathrm{ms}$ \\
$F$ & Walker phasing parameter & $39$
& $[a_{h,2},b_{h,2})$ & GCL window 2 & $[3,3.036)~\mathrm{ms}$ \\
$R_{\oplus}$ & Earth radius & $6371~\mathrm{km}$
& $[0,R)$ & Release-time interval of evaluated packets & $[0,15)~\mathrm{s}$ \\
$\mu$ & Earth gravitational parameter
& $3.986\times10^{14}~\mathrm{m^3/s^2}$
& $\mathcal J$ & Evaluation interval & $[0,16]~\mathrm{s}$ \\
$c$ & Speed of light & $2.998\times10^8~\mathrm{m/s}$
& $W$ & Throughput window length & $500~\mathrm{ms}$ \\
$\ell$ & Packet length & $1500~\mathrm{B}$
& & & \\
\bottomrule
\end{tabular}
\end{table*}

\begin{figure}[t]
\centering
\includegraphics[width=\columnwidth]{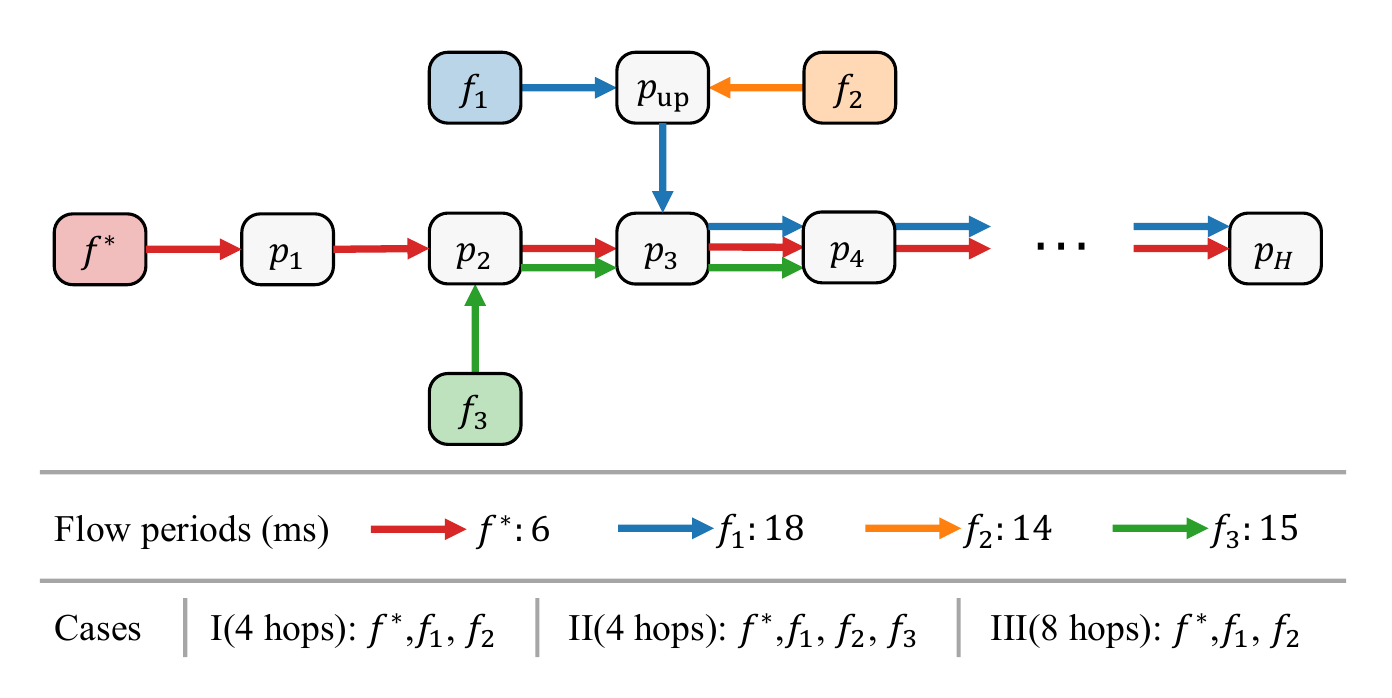}
\caption{Traffic flows and routing configurations.}
\label{fig:traffic-configurations}
\end{figure}

The numerical experiments consider a Walker Delta constellation, with the simulation parameters summarized in Table~\ref{tab:simulation-parameters}. All packets have the same length, and all output ports have the same link rate and use a GCL with two gate-open windows per period. We evaluate the end-to-end delay and jitter bounds of $f^\star$ for packets released during $[0,R)$ and its throughput bound over $[0,W]$. Since the evaluated packets may not reach the destination within $[0,R)$, we use a longer evaluation interval $\mathcal J=[0,E]$ to cover their subsequent transmission and reception, and evaluate the backlog at each output port over the same interval. All traffic sources continue releasing packets throughout $\mathcal J$, as packets released during $[R,E]$ can contribute to port backlog and may compete with target packets still traversing the network.

Fig.~\ref{fig:traffic-configurations} illustrates the traffic flows and routing configurations for the three cases. Case~I uses a four-hop routing path for $f^\star$ and includes flows $f_1$ and $f_2$. Case~II adds $f_3$ while retaining the same four-hop path, whereas Case~III extends the routing path to eight hops with the same traffic configuration as Case~I. To examine the impact of time-varying ISL propagation, all ISLs along the target routing path $\mathcal P_{f^\star}=(p_1,\ldots,p_H)$ are inter-plane links whose propagation delays vary with satellite motion. Flows $f_1$ and $f_2$ release packets directly into the same queue at an upstream output port $p_{\mathrm{up}}$, whose outgoing ISL is an intra-plane link. After traversing this ISL, $f_1$ joins the target path at $p_3$ and continues to $p_H$, competing with $f^\star$ at every output port from $p_3$ to $p_H$. Flow $f_2$ traverses only $p_{\mathrm{up}}$ and does not enter the target path, but can indirectly affect $f^\star$ by influencing the arrival times of $f_1$ at $p_3$. In Case~II, the additional cross flow $f_3$ joins the target path at $p_2$ and continues through $p_3$ to $p_4$, competing with $f^\star$ at all three output ports.

To evaluate the proposed method under different orbital and scheduling configurations, we vary the initial orbital position, the GCL phases of the output ports, and the packet release offsets at the traffic sources for each case. The initial orbital position is specified by $u_0^{(i)}$, which denotes the argument of latitude of the transmitting satellite at $p_1$, with $u_{e_1,0}=u_0^{(i)}$ in \eqref{eq:interplane-delay}. The GCL phases of output ports $p_h$ and $p_{\mathrm{up}}$ are denoted by $\gamma_h^{(j)}$ and $\gamma_{\mathrm{up}}^{(j)}$, respectively, while $\theta_{f_r}^{(k)}$ denotes the packet release offset of flow $f_r$ at its source. These parameters are given by
\begin{equation}
\begin{aligned}
u_0^{(i)}
&=\pi z_i,\\
\gamma_h^{(j)}
&=Tz_{\left((1+(h\bmod4))j+h\right)\bmod5},\\
\gamma_{\mathrm{up}}^{(j)}
&=Tz_j,\\
\theta_{f_r}^{(k)}
&=P_{f_r}z_{\left((r+1)k+r\right)\bmod5},
\end{aligned}
\label{eq:simulation-phase-configurations}
\end{equation}
where $z_m=(2m+1)/10$, $m=0,\ldots,4$, provides five equally spaced normalized values, and $i,j,k\in\{0,\ldots,4\}$, $h=1,\ldots,H$, $r=0,\ldots,3$, with $f_0=f^\star$. The values of $u_0^{(i)}$ sample one period of the inter-plane propagation delay in \eqref{eq:interplane-delay}. The relative orbital phases of the other satellites follow the Walker Delta configuration. All combinations of the three factors are evaluated for Cases I--III, with identical settings for the corresponding output ports and flows across the cases.

For comparison, we consider three network-calculus baselines: STNode~\cite{zhao2018qbv}, FH-TFA~\cite{tabatabaee2023fhtfa}, and STNet~\cite{zhao2021offsets}. All three baselines use the same traffic flows, GCL schedules, routing paths, and target packet set as the proposed method. Their service models account for both gate-open windows in each GCL period and the same number of complete packet transmissions per window, ensuring a common long-term guaranteed service rate at each output port. Variations in ISL propagation delay are also accounted for when updating downstream arrival bounds, and the propagation delay of the final ISL is included in the end-to-end delay bounds. The three baselines are implemented as follows:

\begin{itemize}
    \item STNode uses arrival bounds based on traffic rates and burst sizes, together with GCL service curves, to calculate per-port delay bounds and update downstream arrival bounds. Its service curve is constructed by taking the lower envelope of two periodic service curves, each accounting for both gate-open windows but starting from a different position within the GCL period.

    \item FH-TFA uses periodic staircase arrival curves and updates them along the routing path through link shaping and per-flow delay-jitter bounds. It uses the same TAS service curve as STNode, with finite search intervals determined from linear bounds on arrivals and service.

    \item STNet uses preceding GCL windows and propagation-delay bounds over $\mathcal{J}$ to constrain possible downstream packet arrival times. Delay bounds are calculated over these times by accounting for upstream window shaping and local GCL service.
\end{itemize}

The proposed method is implemented in Python using NumPy, with a subinterval width of $T/64$ for the single-port analysis. The computations are performed on a computer with an Apple M1 processor and 16~GB of memory. The median computation times per configuration, including initialization and result export, are 41.2, 60.0, and 99.7~s for Cases~I--III, respectively. Since network-calculus analysis is primarily intended for offline verification and quantitative comparison of candidate routing and TAS configurations rather than real-time scheduling, these computation times are acceptable for practical use.

\subsection{Validation and Comparison of Performance Bounds}
\label{subsec:performance-bound-validation}

\begin{figure*}[!t]
\centering
\includegraphics[width=0.95\textwidth]{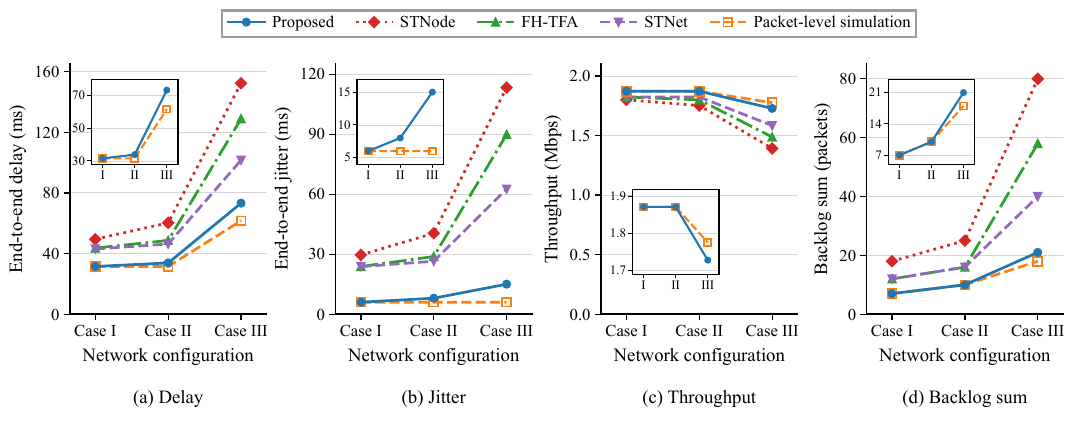}
\caption{Comparison of deterministic performance bounds and packet-level simulation results.}
\label{fig:performance-overview}

\vspace{1em}

\begin{minipage}{\textwidth}
\centering
\footnotesize
\setlength{\tabcolsep}{10pt}
\renewcommand{\arraystretch}{1.1}

\begingroup
\centering
\footnotesize
\setlength{\tabcolsep}{8.5pt}
\renewcommand{\arraystretch}{1.1}

\captionof{table}{Comparison of Deterministic Performance Bounds and Packet-Level Simulation Results}
\label{tab:performance-comparison}

\resizebox{\linewidth}{!}{%
\begin{tabular}{@{}clrrrrrrrcc@{}}
\toprule
\multirow{2}{*}[-0.4ex]{\textbf{Case}}
& \multirow{2}{*}[-0.4ex]{\textbf{Metric}}
& \multicolumn{4}{c}{\textbf{Performance Bounds}}
& \multicolumn{3}{c}{\textbf{Improvement (\%)}}
& \multirow{2}{*}[-0.4ex]{\textbf{Simulation}}
& \multirow{2}{*}[-0.4ex]{\textbf{Gap}} \\
\cmidrule(lr){3-6}
\cmidrule(lr){7-9}
& &
\textbf{STNode} & \textbf{FH-TFA} & \textbf{STNet} & \textbf{Proposed}
& \textbf{STNode} & \textbf{FH-TFA} & \textbf{STNet}
& & \\
\midrule
I & Delay (ms)
& 49.459 & 43.651 & 43.048 & 31.512
& 36.067 & 27.562 & 26.564
& 31.500 & 0.012 \\
& Jitter (ms)
& 29.638 & 23.830 & 23.777 & 6.004
& 79.745 & 74.809 & 74.760
& 5.990 & 0.012 \\
& Throughput (Mbps)
& 1.800 & 1.824 & 1.824 & 1.872
& 4.000 & 2.632 & 2.632
& 1.872 & 0.000 \\
& Backlog Sum (packets)
& 18 & 12 & 12 & 7
& 61.111 & 41.667 & 41.667
& 7 & 0 \\
\midrule
II & Delay (ms)
& 60.293 & 48.663 & 46.306 & 33.836
& 43.772 & 30.177 & 25.986
& 31.512 & 1.988 \\
& Jitter (ms)
& 40.472 & 28.842 & 26.492 & 8.001
& 80.232 & 72.260 & 68.848
& 5.991 & 2.000 \\
& Throughput (Mbps)
& 1.752 & 1.800 & 1.824 & 1.872
& 6.849 & 4.000 & 3.947
& 1.872 & 0.000 \\
& Backlog Sum (packets)
& 25 & 16 & 16 & 10
& 60.000 & 37.500 & 37.500
& 10 & 0 \\
\midrule
III & Delay (ms)
& 152.373 & 128.919 & 101.625 & 73.249
& 51.948 & 43.226 & 28.110
& 61.637 & 10.000 \\
& Jitter (ms)
& 113.286 & 89.833 & 62.542 & 15.013
& 86.749 & 83.289 & 75.627
& 5.991 & 10.012 \\
& Throughput (Mbps)
& 1.392 & 1.488 & 1.584 & 1.728
& 24.138 & 16.129 & 7.576
& 1.776 & 0.048 \\
& Backlog Sum (packets)
& 80 & 58 & 40 & 21
& 73.750 & 63.793 & 47.500
& 18 & 2 \\
\bottomrule
\end{tabular}%
}

\par
\endgroup

\end{minipage}

\end{figure*}

We compare the proposed method with the three baselines and validate the performance bounds using an independent packet-level simulator. In the simulator, packets arriving at the same time are served in order of flow indices, whereas the analytical bounds hold regardless of how ties are resolved. To facilitate an overall comparison of backlog performance, we sum the per-port backlog bounds along the routing path $\mathcal P_{f^\star}$ for each configuration. For backlog, the corresponding simulation result is the sum of the maximum backlogs observed at the individual ports along $\mathcal P_{f^\star}$ during $\mathcal J$. Within each case, we compute the improvement relative to each baseline and the gap to simulation for each configuration before taking their medians. Fig.~\ref{fig:performance-overview} and Table~\ref{tab:performance-comparison} present the performance bounds and simulation results for all four metrics across the three cases, with median values reported over the evaluated configurations. The figure includes inset views for closer comparisons between the proposed bounds and simulation results, while the table further reports the median improvements relative to the baselines and median gaps between the proposed bounds and simulation results.

Across all three cases, the proposed method provides tighter bounds on all four performance metrics than STNode, FH-TFA, and STNet. This advantage becomes more pronounced as the routing path extends from four hops in Case I to eight hops in Case III, where the largest improvements are achieved across all four metrics. The increase is particularly evident for the backlog sum, with improvements relative to STNode, FH-TFA, and STNet rising from 61.1\%, 41.7\%, and 41.7\% in Case I to 73.8\%, 63.8\%, and 47.5\% in Case III, respectively. The jitter results further illustrate the accumulation of conservatism over multiple hops. While the simulated jitter remains approximately 6~ms, the jitter bounds of STNode, FH-TFA, and STNet increase from 29.638, 23.830, and 23.777~ms to 113.286, 89.833, and 62.542~ms, respectively. In contrast, the proposed jitter bound increases only from 6.004 to 15.013~ms, representing reductions of 86.7\%, 83.3\%, and 75.6\% relative to STNode, FH-TFA, and STNet, respectively, in Case III. These results indicate that the baseline analyses become increasingly conservative over longer routing paths, whereas the proposed method substantially limits this increase by accounting for GCL phases and packet ordering.

In addition to outperforming the baselines, the proposed method provides valid bounds that closely match packet-level simulation results across all evaluated configurations. This agreement is particularly close for backlog sum and throughput, with zero median gaps in Cases I and II and gaps of only 2 packets and 0.048~Mbps, respectively, in Case III. By comparison, the delay and jitter gaps increase from approximately 0.012~ms in Case I to approximately 10~ms in Case III. These results show that the proposed method maintains tight backlog and throughput bounds over longer routing paths, although some conservatism remains in the end-to-end delay and jitter bounds.

\begin{figure*}[t]
\centering
\includegraphics[width=0.9\textwidth]{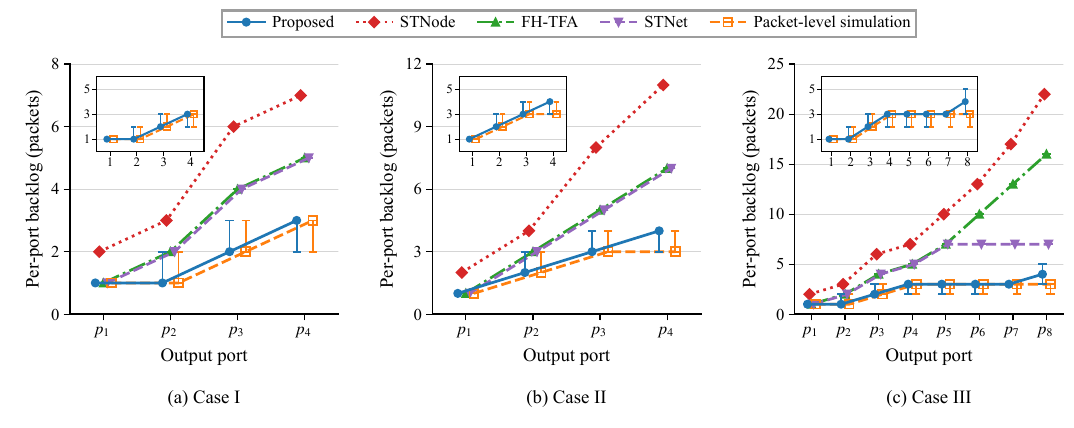}
\caption{Per-port backlog bounds and packet-level simulation results along the routing path.}
\label{fig:backlog-validation}
\end{figure*}

To further evaluate backlog performance, Fig.~\ref{fig:backlog-validation} compares the per-port backlog bounds with the maximum backlogs from packet-level simulation. The figure reports the medians across the evaluated configurations, with error bars indicating the interquartile ranges. The proposed bounds remain close to the simulation results across all three cases, whereas the baselines become increasingly conservative at downstream output ports. This difference is particularly evident in Case III. From $p_4$ to $p_8$, the simulated maximum backlog remains at a median of 3 packets, while the bounds of STNode and FH-TFA increase from 7 to 22 packets and from 5 to 16 packets, respectively. In contrast, the proposed bound increases only from 3 to 4 packets, achieving median reductions of 81.8\% and 75.0\% relative to STNode and FH-TFA, respectively, at $p_8$. The proposed bounds match the simulation results in all configurations at every output port in Case I, the first three ports in Case II, and the first four ports in Case III. These results demonstrate that the proposed method substantially reduces the growth of conservative backlog estimates along longer routing paths.

\subsection{Impact of Time-Varying Propagation on Arrival Bounds}
\label{subsec:propagation-impact}

We use two examples from packet-level simulation to examine how time-varying ISL propagation affects downstream arrival bounds. Consider an arrival interval $[\phi_1(l),\phi_1(r)]$ at $p_2$, which corresponds to a departure interval $[l,r]$ at $p_1$. Following \eqref{eq:isl-arrival-transformation}, the proposed method evaluates $\Gamma_{f^\star,1}^{\mathrm{out}}(l,r-l)$ to bound the number of packets completing serialization at $p_1$ during this interval. For comparison, we also evaluate $\Gamma_{f^\star,1}^{\mathrm{out}}(l,\phi_1(r)-\phi_1(l))$, which uses the arrival interval length instead of the corresponding departure interval length. Both calculations use the same output bound from Section~\ref{subsec:per-flow-output-bounds} and the same time-varying propagation model, differing only in the interval length used to calculate the arrival bound. The actual number of arriving packets is obtained from packet-level simulation.

\begin{table}[t]
\caption{Effects of Time-Varying ISL Propagation on Downstream Arrival Bounds}
\label{tab:propagation-examples}
\centering
\footnotesize
\setlength{\tabcolsep}{3pt}
\renewcommand{\arraystretch}{1.1}
\begin{tabular}{@{}lcc@{}}
\toprule
& \textbf{Case I} & \textbf{Case III} \\
\midrule
Configuration $(i,j,k)$
& $(0,0,4)$ & $(2,2,4)$ \\
Arrival interval start (ms)
& $118.41103820$ & $82.18648343$ \\
Arrival interval end (ms)
& $130.41103580$ & $130.18648563$ \\
Interval length difference (ns)
& $-2.40$ & $+2.20$ \\
\midrule
\multicolumn{3}{@{}l}{Arrival counts (packets)} \\
Packet-level simulation
& $3$ & $8$ \\
Using arrival interval length
& $2$ & $9$ \\
Proposed
& $3$ & $8$ \\
\bottomrule
\end{tabular}
\end{table}

Table~\ref{tab:propagation-examples} presents the results for two arrival intervals of $f^\star$ at $p_2$, selected from configurations $(i,j,k)$ defined in \eqref{eq:simulation-phase-configurations}. The interval length difference is calculated as the arrival interval length minus the corresponding departure interval length. In Case I, the arrival interval is $2.40$~ns shorter than the departure interval. Using the arrival interval length gives a bound of two packets, which is below the three packets observed in simulation. In Case III, the arrival interval is $2.20$~ns longer, and the same calculation gives a bound of nine packets, compared with eight observed in simulation. These examples show that even small differences between the arrival and departure interval lengths can lead to an invalid or overly conservative arrival bound. By using the corresponding departure interval, the proposed method correctly bounds the arrivals in both cases, with results matching the simulation.

\subsection{Contribution of Packet Ordering to Bound Tightness}
\label{subsec:packet-ordering-impact}

\begin{figure*}[!t]
\centering
\includegraphics[width=0.92\textwidth]{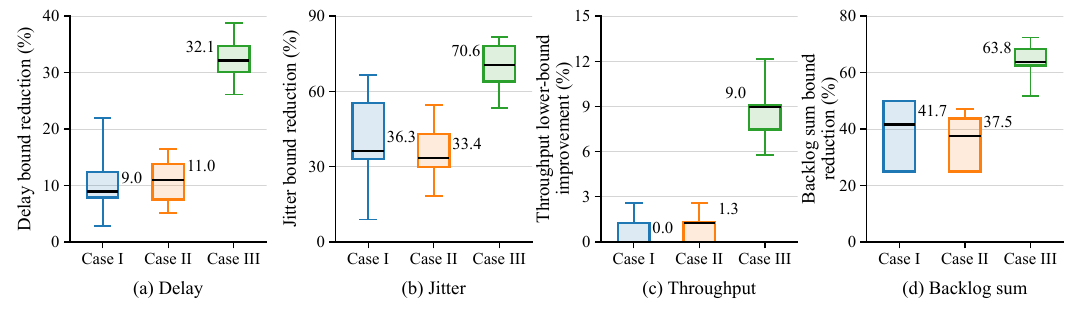}
\caption{Improvements in deterministic performance bounds achieved by packet ordering.}
\label{fig:packet-ordering-reduction}
\end{figure*}

To evaluate the contribution of packet ordering to the tightness of deterministic performance bounds, we compare the proposed method with the same analysis performed without packet ordering. Both analyses use identical network and traffic configurations, GCL service characterization, and time-varying ISL propagation models. 
For each configuration, we calculate the percentage reductions in the delay, jitter, and backlog sum upper bounds and the percentage increase in the throughput lower bound. All percentages are relative to the corresponding bounds obtained without packet ordering. Fig.~\ref{fig:packet-ordering-reduction} shows the distributions of these improvements across the evaluated configurations.

The proposed method generally provides tighter deterministic performance bounds across the evaluated configurations. Throughput improvements are more modest, with a median of zero in Case I despite gains in some configurations. Because the throughput lower bound depends on the number of packets guaranteed to arrive within the throughput window $[0,W]$, tighter delay bounds do not necessarily produce comparable relative improvements in throughput. Compared with the four-hop Case I, the eight-hop Case III achieves greater median improvements in all four metrics, with the jitter improvement increasing from 36.3\% to 70.6\%. These improvements reach 32.1\%, 70.6\%, 9.0\%, and 63.8\% for delay, jitter, throughput, and backlog sum, respectively. Without packet ordering, conservative estimates of competing packets at each output port can affect the bounds at downstream ports, allowing conservatism to accumulate over multiple hops. By excluding packets that cannot delay the packet under analysis, packet ordering reduces this accumulation and tightens deterministic performance bounds, with greater improvements along the longer routing path in Case III.

\subsection{Parameter Sensitivity Analysis}
\label{subsec:parameter-sensitivity}

We further examine the sensitivity of deterministic performance bounds to GCL window length and cross-traffic load under the network and traffic configuration of Case~II. For each parameter value, we evaluate all combinations of initial orbital position, GCL phases, and packet release offsets corresponding to the indices $i$, $j$, and $k$ defined in \eqref{eq:simulation-phase-configurations} and report the medians of the performance bounds and packet-level simulation results.

\begin{figure*}[!t]
\centering
\includegraphics[width=0.95\textwidth]{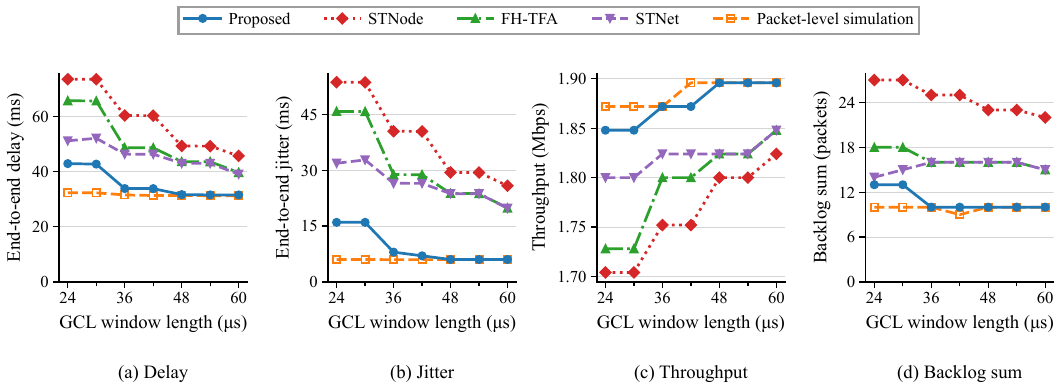}
\caption{Impact of GCL window length on deterministic performance bounds.}
\label{fig:gcl-window-sensitivity}
\end{figure*}

\begin{figure*}[!t]
\centering
\includegraphics[width=0.95\textwidth]{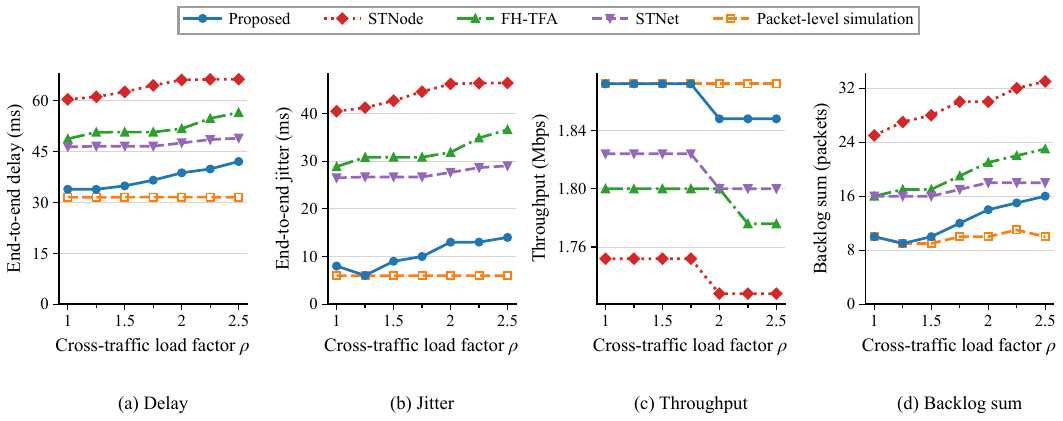}
\caption{Impact of cross-traffic load on deterministic performance bounds.}
\label{fig:cross-traffic-sensitivity}
\end{figure*}

Fig.~\ref{fig:gcl-window-sensitivity} shows the performance bounds and packet-level simulation results as the common length $w$ of the two GCL gate-open windows increases from 24 to 60~$\mu$s in steps of 6~$\mu$s. The GCL period remains fixed at $T=8$~ms, the windows are given by $[0,w)$ and $[3~\mathrm{ms},3~\mathrm{ms}+w)$, and all other settings remain unchanged from Case~II. Over this range, the median delay, jitter, and backlog sum bounds of the proposed method decrease by 26.5\%, 62.4\%, and 23.1\%, respectively, while the median throughput lower bound increases by 2.6\%. The largest reductions in delay and jitter occur when the window length increases from 30 to 36~$\mu$s, with the corresponding bounds decreasing by 20.8\% and 50.0\%, respectively. This is because each packet requires 12~$\mu$s for serialization, and increasing the window length from 30 to 36~$\mu$s allows three rather than two packets to complete transmission within each window, reducing the need to wait for subsequent GCL windows. Beyond 48~$\mu$s, further widening provides smaller improvements, and the medians of all four bounds remain unchanged between 54 and 60~$\mu$s. At 60~$\mu$s, the median delay and jitter bounds of the proposed method are 31.524 and 6.016~ms, compared with 31.248 and 5.991~ms in simulation. In contrast, the delay and jitter bounds of the three baselines remain substantially higher than the simulation results even at the largest evaluated window length. By accounting for the timing of packet arrivals relative to GCL windows and excluding packets that cannot delay the target packet, the proposed method reduces the overestimation of packet waiting times. Overall, widening GCL windows improves deterministic performance bounds primarily by increasing the number of packets that can be transmitted within each window, while the benefits diminish as the service capacity increases relative to the traffic load.

Fig.~\ref{fig:cross-traffic-sensitivity} presents the results under different cross-traffic loads. We retain the original GCL settings and the period of $f^\star$, while adjusting the periods of the three cross flows according to $P_{f_r}(\rho)=P_{f_r}^{(0)}/\rho$ for $r=1,2,3$, where $P_{f_r}^{(0)}$ denotes the original flow period given in Fig.~\ref{fig:traffic-configurations}. The load factor $\rho$ increases from 1 to 2.5 in steps of 0.25, and the packet release offsets of the cross flows are recalculated according to \eqref{eq:simulation-phase-configurations}. Over this range, the median delay, jitter, and backlog sum bounds of the proposed method increase by 24.1\%, 75.0\%, and 60.0\%, respectively. However, the median delay observed in simulation remains between 31.500 and 31.512~ms, while the median gap between the proposed delay bound and simulation increases from 1.988 to 9.000~ms. The gap increases because higher cross-traffic arrival bounds increase the number of potentially competing packets at shared output ports. When the number of packets to be transmitted exceeds the capacity of the current GCL window, the completion time bound extends to a subsequent window, and the resulting conservatism propagates downstream. In contrast, the throughput lower bound decreases by only 1.3\%, as the median number of target packets guaranteed to reach the destination within the throughput window decreases from 78 to 77, compared with 78 in simulation. Despite the increased conservatism, the proposed method maintains substantially tighter bounds than the three baselines. At the highest evaluated load ($\rho=2.5$), it achieves median reductions of 36.4\%, 25.4\%, and 13.7\% in delay bounds and 69.8\%, 61.7\%, and 51.5\% in jitter bounds relative to STNode, FH-TFA, and STNet, respectively.

In both sensitivity experiments, the proposed method provides tighter bounds on all four performance metrics than all three baselines across all evaluated parameter values. Moreover, the benefits of widening GCL windows depend on whether it increases the per-window packet transmission capacity, while heavier cross-traffic loads generally increase the conservatism of multihop analysis.

\section{Conclusion}
\label{sec:conclusion}

This paper established an absolute-time network calculus framework for deriving deterministic end-to-end performance bounds for TAS in LEO satellite networks. In this framework, we characterize packet arrivals and GCL service over absolute time intervals and account for time-varying ISL propagation. Packet ordering is further used to exclude packets that cannot delay the target packet, thereby reducing conservatism in multihop analysis. Numerical results show that the proposed method provides tighter bounds than three network-calculus baselines, with the resulting bounds closely matching packet-level simulation results. Beyond verifying whether a given routing and TAS configuration satisfies the performance requirements of TS services, the framework can also support TAS configuration design. Comparing performance bounds across different configurations provides quantitative guidance for selecting GCL parameters.

\bibliographystyle{IEEEtran}
\bibliography{references}

@article{zhu2025beam,
  author = {Jianfeng Zhu and others},
  title = {Beam Management in Low Earth Orbit Satellite Communication With Handover Frequency Control and Satellite-Terrestrial Spectrum Sharing},
  journal = {IEEE Transactions on Communications},
  volume = {73},
  number = {7},
  pages = {5247--5263},
  month = jul,
  year = {2025},
  note = {doi: 10.1109/TCOMM.2024.3516479},
  doi = {10.1109/TCOMM.2024.3516479}
}

@article{feng2024routing,
  author = {Xin'ao Feng and others},
  title = {Distributed Satellite-Terrestrial Cooperative Routing Strategy Based on Minimum Hop-Count Analysis in Mega {LEO} Satellite Constellation},
  journal = {IEEE Transactions on Mobile Computing},
  volume = {23},
  number = {11},
  pages = {10678--10693},
  month = nov,
  year = {2024},
  note = {doi: 10.1109/TMC.2024.3380891},
  doi = {10.1109/TMC.2024.3380891}
}

@article{li2026safvin,
  author = {Dongbo Li and Bo Yin and Daohua Yan and Jie Liu and Guoliang Xing and Zhijun Li},
  title = {{SAFVIN}: Edge Intelligence for Satellite and Autonomous Farm Vehicle Integrated Networks},
  journal = {IEEE Transactions on Mobile Computing},
  volume = {25},
  number = {7},
  pages = {11305--11320},
  month = jul,
  year = {2026},
  note = {doi: 10.1109/TMC.2026.3665486},
  doi = {10.1109/TMC.2026.3665486}
}

@article{sun2026handover,
  author = {Yaohua Sun and others},
  title = {A User-Grouping-Based Network-Orchestrated Inter-Satellite Handover Approach in Heterogeneous Satellite Networks},
  journal = {IEEE Internet of Things Journal},
  volume = {13},
  number = {23},
  pages = {60938--60956},
  month = dec,
  year = {2026},
  note = {doi: 10.1109/JIOT.2026.3740906},
  doi = {10.1109/JIOT.2026.3740906}
}

@article{lobello2019perspective,
  author = {Lucia {Lo Bello} and Wilfried Steiner},
  title = {A Perspective on {IEEE} Time-Sensitive Networking for Industrial Communication and Automation Systems},
  journal = {Proceedings of the IEEE},
  volume = {107},
  number = {6},
  pages = {1094--1120},
  month = jun,
  year = {2019},
  note = {doi: 10.1109/JPROC.2019.2905334},
  doi = {10.1109/JPROC.2019.2905334}
}

@misc{zein2025spaceprofile,
  author = {Nader Zein},
  title = {{TSN} Profile for Terrestrial Space and Multi Orbits Satellite to Satellite Communications},
  howpublished = {IEEE 802.1 Working Group contribution},
  year = {2025},
  month = jul,
  url = {https://www.ieee802.org/1/files/public/docs2025/new-zein-TSN-Profile-for-Terrestrial-Space-and-Multi-Orbits-Satellite-to-Satellite-Communications-0725.pdf}
}

@inproceedings{mohan2024starlink,
  author = {Nitinder Mohan and Andrew E. Ferguson and Hendrik Cech and Rohan Bose and Prakita Rayyan Renatin and Mahesh K. Marina and J{\"o}rg Ott},
  title = {A Multifaceted Look at {Starlink} Performance},
  booktitle = {Proc. ACM Web Conference (WWW)},
  pages = {2723--2734},
  year = {2024},
  note = {doi: 10.1145/3589334.3645328},
  doi = {10.1145/3589334.3645328}
}

@standard{ieee8021q2022,
  title = {{IEEE} Standard for Local and Metropolitan Area Networks---Bridges and Bridged Networks},
  organization = {IEEE},
  number = {802.1Q-2022},
  year = {2022},
  note = {doi: 10.1109/IEEESTD.2022.10004498},
  doi = {10.1109/IEEESTD.2022.10004498}
}

@techreport{ieee80224tsn2019,
  author = {Ruben Salazar and Tim Godfrey and Ludwig Winkel and Norm Finn and Clint Powell and Ben Rolfe and Maik Seewald},
  title = {Utility Applications of Time Sensitive Networking White Paper},
  institution = {IEEE 802.24 Vertical Applications Technical Advisory Group},
  type = {White Paper},
  year = {2019}
}

@article{zhao2022shapers,
  author = {Luxi Zhao and Paul Pop and Sebastian Steinhorst},
  title = {Quantitative Performance Comparison of Various Traffic Shapers in Time-Sensitive Networking},
  journal = {IEEE Transactions on Network and Service Management},
  volume = {19},
  number = {3},
  pages = {2899--2928},
  month = sep,
  year = {2022},
  note = {doi: 10.1109/TNSM.2022.3180160},
  doi = {10.1109/TNSM.2022.3180160}
}

@inproceedings{craciunas2016scheduling,
  author = {Silviu S. Craciunas and Ramon {Serna Oliver} and Martin Chmel{\'i}k and Wilfried Steiner},
  title = {Scheduling Real-Time Communication in {IEEE 802.1Qbv} Time Sensitive Networks},
  booktitle = {Proceedings of the 24th International Conference on Real-Time Networks and Systems},
  pages = {183--192},
  year = {2016},
  note = {doi: 10.1145/2997465.2997470},
  doi = {10.1145/2997465.2997470}
}

@techreport{rfc9320,
  author = {Norman Finn and Jean-Yves {Le Boudec} and Ehsan Mohammadpour and Jiayi Zhang and Balazs Varga},
  title = {Deterministic Networking ({DetNet}) Bounded Latency},
  institution = {Internet Engineering Task Force (IETF)},
  type = {RFC},
  number = {9320},
  year = {2022},
  month = nov,
  note = {doi: 10.17487/RFC9320},
  doi = {10.17487/RFC9320}
}

@misc{donnelly2022nasa,
  author = {Rob Donnelly},
  title = {{TSN} Time Synchronization: {NASA}'s Use Cases and Needs},
  howpublished = {IEEE P802.1DP/SAE AS6675 Ad Hoc contribution},
  year = {2022},
  month = may,
  url = {https://www.ieee802.org/1/files/public/docs2022/dp-donnelly-NASA-needs-0522-v00.pdf}
}

@misc{pasquier2015afdx,
  author = {Bruno Pasquier and Stefan Schneele},
  title = {Avionics Full Duplex Ethernet and the Time Sensitive Networking Standard},
  howpublished = {IEEE 802.1 TSN Working Group contribution},
  year = {2015},
  month = may,
  url = {https://www.ieee802.org/1/files/public/docs2015/TSN-Schneele-AFDX-0515-v01.pdf}
}

@book{leboudec2001networkcalculus,
  author = {Jean-Yves {Le Boudec} and Patrick Thiran},
  title = {Network Calculus: A Theory of Deterministic Queuing Systems for the Internet},
  series = {Lecture Notes in Computer Science},
  volume = {2050},
  publisher = {Springer},
  address = {Berlin, Heidelberg},
  year = {2001},
  note = {doi: 10.1007/3-540-45318-0},
  doi = {10.1007/3-540-45318-0}
}

@article{jiang2023qoe,
  author = {Dingde Jiang and Feng Wang and Zhihan Lv and Shahid Mumtaz and Saba Al-Rubaye and Antonios Tsourdos and Octavia Dobre},
  title = {{QoE}-Aware Efficient Content Distribution Scheme for Satellite-Terrestrial Networks},
  journal = {IEEE Transactions on Mobile Computing},
  volume = {22},
  number = {1},
  pages = {443--458},
  month = jan,
  year = {2023},
  note = {doi: 10.1109/TMC.2021.3074917},
  doi = {10.1109/TMC.2021.3074917}
}

@article{zhao2018qbv,
  author = {Luxi Zhao and Paul Pop and Silviu S. Craciunas},
  title = {Worst-Case Latency Analysis for {IEEE 802.1Qbv} Time Sensitive Networks Using Network Calculus},
  journal = {IEEE Access},
  volume = {6},
  pages = {41803--41815},
  year = {2018},
  note = {doi: 10.1109/ACCESS.2018.2858767},
  doi = {10.1109/ACCESS.2018.2858767}
}

@article{zhao2021offsets,
  author = {Luxi Zhao and Paul Pop and Zijie Gong and Bingwu Fang},
  title = {Improving Latency Analysis for Flexible Window-Based {GCL} Scheduling in {TSN} Networks by Integration of Consecutive Nodes Offsets},
  journal = {IEEE Internet of Things Journal},
  volume = {8},
  number = {7},
  pages = {5574--5584},
  month = apr,
  year = {2021},
  note = {doi: 10.1109/JIOT.2020.3031932},
  doi = {10.1109/JIOT.2020.3031932}
}

@article{hu2025upperbound,
  author = {Wenxue Hu and Lei Sun and Jianquan Wang and Wei Chen and Wei Li},
  title = {Upper-Bound Latency Analysis for {TSN} With Improved Network Calculus-Based Model},
  journal = {Electronics Letters},
  volume = {61},
  note = {Art. no. e70302, doi: 10.1049/ell2.70302},
  year = {2025},
  doi = {10.1049/ell2.70302}
}

@article{daniel2013timevarying,
  author = {M{\'a}bia Daniel-Cavalcante and Rafael Santos-Mendes},
  title = {Modular Methodology for the Network Calculus in a Time-Varying Context},
  journal = {IEEE Transactions on Information Theory},
  volume = {59},
  number = {10},
  pages = {6342--6356},
  month = oct,
  year = {2013},
  note = {doi: 10.1109/TIT.2013.2272870},
  doi = {10.1109/TIT.2013.2272870}
}

@inproceedings{falk2019intermittence,
  author = {Jonathan Falk and Frank D{\"u}rr and Kurt Rothermel},
  title = {Modeling Time-Triggered Service Intermittence in Network Calculus},
  booktitle = {Proc. 27th International Conference on Real-Time Networks and Systems (RTNS)},
  pages = {90--100},
  year = {2019},
  note = {doi: 10.1145/3356401.3356411},
  doi = {10.1145/3356401.3356411}
}

@article{wang2025leodelay,
  author = {Huizi Wang and Lei Sun and Jianquan Wang and Shangjing Lin and Zhuoqun Wang and Kewen Sun},
  title = {Research on Delay Upper Bound Analysis Based on Network Calculus for {LEO} Satellite Network},
  journal = {Journal on Communications},
  volume = {46},
  number = {4},
  pages = {80--90},
  month = apr,
  year = {2025},
  note = {In Chinese, doi: 10.11959/j.issn.1000-436x.2025074},
  doi = {10.11959/j.issn.1000-436x.2025074}
}

@article{zhang2025martingale,
  author = {Qingqing Zhang and Yan Zhu and Di Zhou and Yan Dong and Shun Guo and Weidang Lu and Zhu Han},
  title = {Martingale Theory-Based Delay Bound Analysis for Multi-Hop Heterogeneous Satellite Networks},
  journal = {IEEE Transactions on Communications},
  volume = {73},
  number = {7},
  pages = {5198--5213},
  month = jul,
  year = {2025},
  note = {doi: 10.1109/TCOMM.2024.3511710},
  doi = {10.1109/TCOMM.2024.3511710}
}

@article{thomas2020clocks,
  author = {Ludovic Thomas and Jean-Yves {Le Boudec}},
  title = {On Time Synchronization Issues in Time-Sensitive Networks With Regulators and Nonideal Clocks},
  journal = {Proceedings of the ACM on Measurement and Analysis of Computing Systems},
  volume = {4},
  number = {2},
  pages = {1--41},
  note = {Art. no. 27, doi: 10.1145/3392145},
  month = jun,
  year = {2020},
  doi = {10.1145/3392145}
}

@article{chen2026martingale,
  author = {Mengmei Chen and Liu Chao and Xiaorong Zhu},
  title = {An End-to-End Delay Analysis Model of Multi-Traffic for Multi-Hop {LEO} Satellite Network Based on Martingale Theory},
  journal = {IEEE Transactions on Network Science and Engineering},
  volume = {13},
  pages = {5158--5175},
  year = {2026},
  note = {doi: 10.1109/TNSE.2025.3642552},
  doi = {10.1109/TNSE.2025.3642552}
}

@inproceedings{tabatabaee2023fhtfa,
  author = {Seyed Mohammadhossein Tabatabaee and Marc Boyer and Jean-Yves {Le Boudec} and J{\"o}rn Migge},
  title = {Efficient and Accurate Handling of Periodic Flows in Time-Sensitive Networks},
  booktitle = {Proc. IEEE 29th Real-Time and Embedded Technology and Applications Symposium (RTAS)},
  pages = {303--315},
  year = {2023},
  note = {doi: 10.1109/RTAS58335.2023.00031},
  doi = {10.1109/RTAS58335.2023.00031}
}

@article{bouillard2015exact,
  author = {Anne Bouillard and Giovanni Stea},
  title = {Exact Worst-Case Delay in {FIFO}-Multiplexing Feed-Forward Networks},
  journal = {IEEE/ACM Transactions on Networking},
  volume = {23},
  number = {5},
  pages = {1387--1400},
  month = oct,
  year = {2015},
  note = {doi: 10.1109/TNET.2014.2332071},
  doi = {10.1109/TNET.2014.2332071}
}

@article{geyer2023flow,
  author = {Fabien Geyer and Alexander Scheffler and Steffen Bondorf},
  title = {Network Calculus With Flow Prolongation---A Feedforward {FIFO} Analysis Enabled by {ML}},
  journal = {IEEE Transactions on Computers},
  volume = {72},
  number = {1},
  pages = {97--110},
  month = jan,
  year = {2023},
  note = {doi: 10.1109/TC.2022.3204225},
  doi = {10.1109/TC.2022.3204225}
}

@article{yuan2024placement,
  author = {Shuo Yuan and others},
  title = {Joint Network Function Placement and Routing Optimization in Dynamic Software-Defined Satellite-Terrestrial Integrated Networks},
  journal = {IEEE Transactions on Wireless Communications},
  volume = {23},
  number = {5},
  pages = {5172--5186},
  month = may,
  year = {2024},
  note = {doi: 10.1109/TWC.2023.3324729},
  doi = {10.1109/TWC.2023.3324729}
}

@article{zhu2024timing,
  author = {Jianfeng Zhu and others},
  title = {Timing Advance Estimation in Low Earth Orbit Satellite Networks},
  journal = {IEEE Transactions on Vehicular Technology},
  volume = {73},
  number = {3},
  pages = {4366--4382},
  month = mar,
  year = {2024},
  note = {doi: 10.1109/TVT.2023.3325328},
  doi = {10.1109/TVT.2023.3325328}
}

@article{feng2026jrs,
  author = {Xin'ao Feng and others},
  title = {Joint Multipath Routing and Traffic Scheduling for Time-Sensitive Services in Mega {LEO} Constellation},
  journal = {IEEE Transactions on Mobile Computing},
  month = "early access, " # nov # "~2,",
  year = {2026},
  note = {doi: 10.1109/TMC.2026.3749726},
  doi = {10.1109/TMC.2026.3749726}
}

@article{lu2026lecmor,
  author = {Mai-Kao Lu and Yaowei Xu and Yunxiang Chen and Rui Li and Yutong Liu and Caiqian Zhang and Baojun Lin and Wenbin Gong and Linghe Kong and Guihai Chen},
  title = {{LeCMOR}: A Learning-Enhanced Centralized Routing Strategy for Congestion Mitigation in Multi-Layer Laser Space Networks},
  journal = {IEEE Transactions on Mobile Computing},
  volume = {25},
  number = {11},
  pages = {19874--19889},
  month = nov,
  year = {2026},
  note = {doi: 10.1109/TMC.2026.3709185},
  doi = {10.1109/TMC.2026.3709185}
}

@article{ren2026meta,
  author = {Yuzheng Ren and Yangyang Hu and Haijun Zhang and Yuxin Luo and Jingyi Ma and Fei Richard Yu and Xueyan Cao},
  title = {Meta-Optimization for Discovering Adaptive Routing Rules in Dynamic Heterogeneous {LEO} Satellite Networks},
  journal = {IEEE Transactions on Mobile Computing},
  volume = {25},
  number = {12},
  pages = {23666--23682},
  month = dec,
  year = {2026},
  note = {doi: 10.1109/TMC.2026.3730035},
  doi = {10.1109/TMC.2026.3730035}
}

@article{chen2024shortest,
  author = {Quan Chen and Lei Yang and Yong Zhao and Yi Wang and Haibo Zhou and Xiaoqian Chen},
  title = {Shortest Path in {LEO} Satellite Constellation Networks: An Explicit Analytic Approach},
  journal = {IEEE Journal on Selected Areas in Communications},
  volume = {42},
  number = {5},
  pages = {1175--1187},
  month = may,
  year = {2024},
  note = {doi: 10.1109/JSAC.2024.3365873},
  doi = {10.1109/JSAC.2024.3365873}
}

\end{document}